\documentclass[hyper,11pt,paper]{article}
\usepackage{jheppub} 

\usepackage[T1]{fontenc} 
\usepackage{tabu}
\usepackage{graphicx,amsmath,amssymb,multirow,array,bm,mathrsfs}
\usepackage[mathscr]{eucal}
\usepackage{upgreek}
\usepackage{amsthm}
\usepackage{arydshln}
\usepackage{color}
\usepackage{rotating}
\usepackage{tikz}

\newcommand{\mC}{\mathcal{C}}
\newcommand{\fH}{\mathbb{H}}
\newcommand{\fZ}{\mathbb{Z}}

\newcommand{\mO}{\mathscr{O}}

\newcommand{\mD}{\mathcal{D}}
\newcommand{\mZ}{\mathrm{Z}}
\newcommand{\tD}{\tilde{D}}
\newcommand{\dee}{d_1}
\newcommand{\eps}{\varepsilon}

\newcommand{\cindex}{{\mathfrak Z}}
\newcommand{\ON}{\Omega_N}
\newcommand{\Oinf}{\Omega_\infty}
\newcommand{\ua}{\uparrow}
\newcommand{\da}{\downarrow}
\newcommand{\no}{\circ}
\newcommand{\yes}{\bullet}

\def\CNfold{{\cal C}^{\otimes N}}
\def\Cperm{{\cal C}_{N,\Omega}}
\def\Csym{{\cal C}_{N,S}}
\def\Ccyc{{\cal C}_{N,{\mathbb Z}}}
\def\sym{N,S}
\def\symq{q,S}
\def\symp{p,S}
\def\cyc{N,{\mathbb Z}}
\def\perm{N,\Omega}
\def\cc{\textbf c}

\def\bq{{\bar q}}
\def\bc{{\bar c}}
\def\bh{{\bar h}}
\def\bn{{\bar n}}
\def\btau{{\bar \tau}}
\def\bchi{\bar \chi}
\def\cn{\frac{c}{24}}
\def\bcn{\frac{\bc}{24}}

\newtheorem{theorem}{Theorem}

\definecolor{rust}{rgb}{0.8,0.2,0.2}
\definecolor{green}{rgb}{0,0.3,0.2}
\definecolor{Orange}{rgb}{1,.4,0}

\def\AdS#1{AdS$_{#1}$}
\def\CFT#1{CFT$_{#1}$}

\title{Permutation orbifolds and holography}

\author[]{Felix M. Haehl}
\author[]{\!, Mukund Rangamani}
\affiliation[]{Centre for Particle Theory \& Department of Mathematical Sciences,\\
                     Durham University, South Road, Durham DH1 3LE, UK.}
\emailAdd{f.m.haehl@gmail.com}
\emailAdd{mukund.rangamani@durham.ac.uk}

\abstract{ 
Two dimensional conformal field theories with large central charge and a  sparse low-lying spectrum are expected to admit a classical string holographic dual. We construct a large class of such theories employing permutation orbifold technology. In particular, we describe the group theoretic constraints on permutation groups to ensure a (stringy) holographic CFT. The primary result we uncover is that in order for the degeneracy of states to be finite in the large central charge limit, the groups of interest  are the so-called oligomorphic permutation groups. Further requiring that the low-lying spectrum be sparse enough puts a bound on the number of orbits of these groups  (on finite element subsets).   Along the way we also study familiar cyclic and symmetric orbifolds to build intuition. We also demonstrate how holographic spectral properties are tied to the geometry of covering spaces for permutation orbifolds.
}

\begin{document} 
	\begin{flushright} \small{DCPT-14/69} \end{flushright}
	
\maketitle
\flushbottom

\section{Introduction}
\label{sec:introduction}

The holographic gauge/gravity correspondence posits a relation between quantum field theories and a class of gravitational theories. In the well understood cases the gravitational dynamics arises in a certain strongly coupled, large central charge limit of the field theory, and typically is given by the familiar Einstein gravity (perhaps coupled to matter) in asymptotically AdS spacetimes. However we now appreciate that there are more exotic examples wherein the gravitational dynamics is part of a higher spin theory, cf., \cite{Gaberdiel:2012uj,Giombi:2012ms}. 

Given a quantum field theory one would like to write down the conditions required for it to admit a holographic dual. Ideally, this condition would delineate the class of two derivative gravity theories from their more exotic cousins, for in the former we have a much cleaner understanding of spacetime geometry. At a heuristic level familiar gravitational dynamics arises when the field theories in question have  `matrix-like' degrees of freedom, while higher-spin dynamics is associated with `vector-like' degrees of freedom. A-priori, while one can only argue that theories with matrix-like degrees of freedom are associated typically with stringy duals, in a suitable strong coupling, large central charge (planar) limit, this  string dynamics truncates to classical gravitational dynamics.

This statement was made precise by \cite{Heemskerk:2009pn} who argued that in order for the bulk gravitational theory to be local on length scales smaller than the bulk AdS scale, it must not only have a large number of degrees of freedom, but also have a sparse low lying spectrum of excitations. One can understand this from familiar examples: in the planar $N\to \infty$, strongly coupled limit of  ${\cal N} =4$ SYM (with gauge group $SU(N)$) the stringy excitations with spin $s >2$ get infinitely heavy \cite{Gubser:1998bc}. 
Similar conclusions were also reached in the context of low dimensional CFTs by \cite{ElShowk:2011ag} where, using intuition from the \AdS{3}/\CFT{2} correspondence, it was argued that the sparse low lying spectrum of (super)graviton type excitations should be complemented by a large degeneracy of states above a gap (corresponding to black hole microstates). 

However, it can be argued that in general the two criteria (i) large central charge and (ii) sparse low-lying spectrum are by themselves insufficient to distinguish classical gravity duals from classical string duals. A simple case in point is the symmetric product orbifold in two dimensions. The familiar D1-D5 brane system in the decoupling limit gives rise to string theory on \AdS{3} $\times {\bf S}^3 \times K3$ with the world-volume dynamics reducing to the two dimensional CFT with target space $(K3)^{Q_1\,Q_5}/S_{Q_1Q_5}$. However, the CFT at the free orbifold point is singular and presumably only corresponds to the tensionless limit of the dual string theory. The supergravity limit arises by deforming away from the free point by a marginal operator.

Despite this important distinction, one can make a case for generic symmetric product orbifold CFTs (even without supersymmetry) to display properties which one can understand from the dual classical gravitational description \cite{Keller:2011xi}. In particular, once the aforementioned criteria are satisfied, it is possible to show that the canonical free energy of the CFT (which encodes the spectral density of states)  undergoes a sharp phase transition at an ${\cal O}(1)$ temperature, in fact at  $T_c = \frac{1}{2\pi}$. In gravity this is a manifestation of the Hawking-Page transition between the thermal \AdS{3} and BTZ geometries. On the CFT side the result is a consequence of  modular invariance of two dimensional CFTs, as it relates the low and high energy density of states, and thus can be used to give a very precise characterization of the sparseness. This analysis was carried out recently in the beautiful work \cite{Hartman:2014oaa} who gave precise bounds on the growth rate of the density of states.\footnote{ Constraints from modular invariance on the spectrum of two dimensional CFTs has been much explored since the original work of \cite{Hellerman:2009bu}; see \cite{Friedan:2013cba,Qualls:2013eha} for other salient results in this vein.}

So how does one understand the lack of distinction between classical (tensionless) string and gravity descriptions based on the spectral information? We believe the situation is analogous to the thermal behaviour of gauge theories on compact spaces. Consider a four dimensional ${\cal N}=4$ SYM on ${\bf S}^3 \times {\mathbb R}$ which undergoes a finite temperature  first order phase transition (akin to a Hagedorn transition) at vanishing coupling ($\lambda =0$) \cite{Sundborg:1999ue,Aharony:2003sx}. While this is qualitatively similar to the Hawking-Page transition, one expects that the free theory phase transition resolves at non-vanishing coupling to two distinct transitions.\footnote{ This is indeed what happens in the non-supersymmetric pure glue theory \cite{Aharony:2005bq}; the precise behaviour for ${\cal N} =4 $ SYM remains an open question.} The lower first order transition has been conjectured to interpolate toward the Hawking-Page transition in the strong coupling limit, while the higher third order transition is thought to interpolate toward the stringy Hagedorn transition (which is hierarchically separated from the gravitational transition). The former occurs at $T \sim {\cal O}(1)$ while the latter occurs at 
$T \sim {\cal O}(\lambda^\frac{1}{4})$ in the $\lambda \gg 1$ limit. 

While this conflation of the two transitions in the weak coupling limit is unfortunate, it  should be borne in mind that it still provides a clear distinction between the vectorial and matrix-like theories. In the former, even in the large central charge limit, there is an abundance of states at low energies which tends to wash out the finite temperature phase transition (more precisely it pushes into a highly quantum regime). This was established for vector models in 2+1 dimensions in \cite{Shenker:2011zf} (see also \cite{Banerjee:2012gh}) and one expects something similar the context of two dimensional minimal models  \cite{Banerjee:2012aj,Gaberdiel:2013cca}. 

With these caveats in mind, let us turn to the main question that drove this investigation. In the space of two dimensional CFTs is there a natural characterization of the class of matrix like theories which have a hope of giving rise to stringy holographic duals (which furthermore, in particular corners of moduli space, might reduce to gravitational dynamics on an asymptotically \AdS{3} spacetime)? Let us call this family of theories {\em stringy holographic CFTs}; they will be distinguished by satisfying the two criteria set out above.

 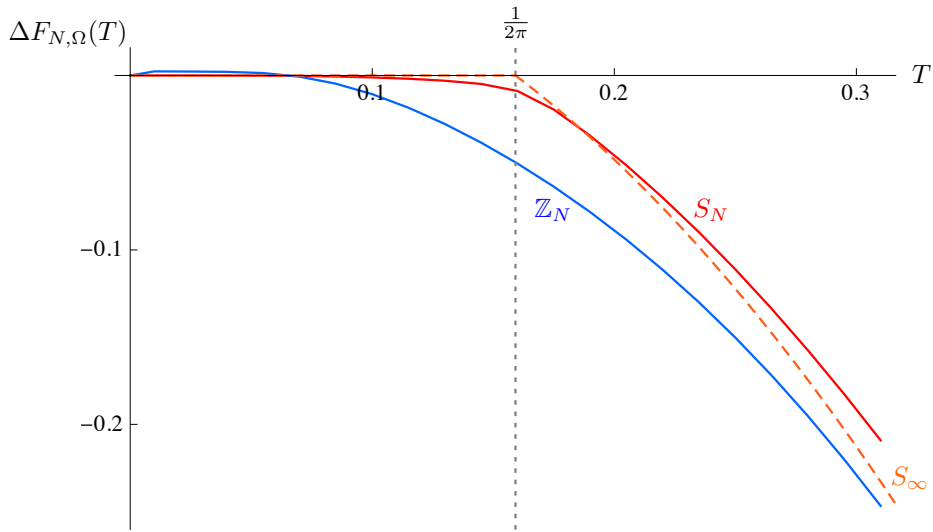
\begin{figure}
 \small
 \centerline{\includegraphics[width=.7\textwidth]{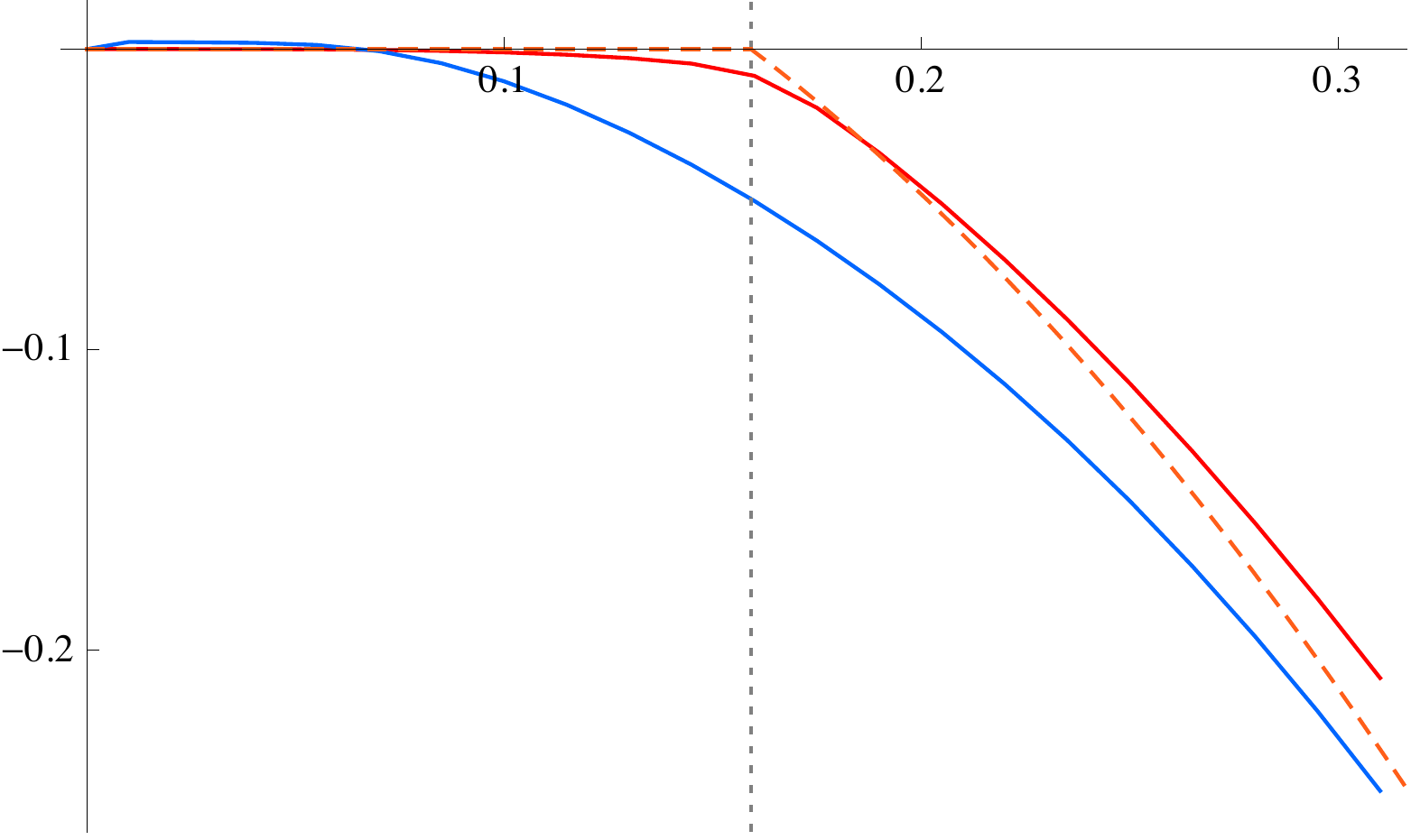}}
  \setlength{\unitlength}{0.1\columnwidth}
 \begin{picture}(0.3,0.4)(0,0)
 \put(1.4,4.7){\makebox(0,0){$\Delta F_{\perm}(T)$}}
 \put(8.7,4.36){\makebox(0,0){$T$}}
 \put(5.23,4.8){\makebox(0,0){$\tfrac{1}{2\pi}$}}
 \put(8.6,.9){\makebox(0,0){{\color{Orange}$S_\infty$}}}
 \put(6.9,3.2){\makebox(0,0){{\color{red}$S_N$}}}
 \put(5.55,3.2){\makebox(0,0){{\color{blue}$\fZ_N$}}}
 \end{picture}
 \caption{Plot of the vacuum subtracted rescaled free energies defined in \eqref{eq:deltaF} of orbifold CFTs $\Cperm$ with respect to a cyclic group $\fZ_N$ (blue line) and a symmetric group $S_N$ (red line) for $N=103$. 
 ${\cal C}$ is taken to be the free boson CFT at unit radius. The dashed (orange) line shows the limit of the symmetric orbifold as $N\rightarrow \infty$. Any other CFT with a holographic dual has the same universal large central charge asymptotics.  The vertical line is $T =\frac{1}{2\pi}$, drawn to guide the eye for the critical temperature, and corresponds to a square torus. 
 }
 \label{fig:plot}
 \end{figure}

In the present paper we undertake the exercise of constructing a large class of string holographic CFTs in two  dimensions. The basic tool we will use is the fact that large central charge theories can be naturally constructed by taking tensor product of some (small central charge) CFT and using orbifold technology to ensure sparseness in the spectrum. Let us state the precise problem we wish to tackle. Consider a CFT ${\cal C}$ with central charge $c$. We will assume that this theory has a gap with the lowest primary ${\cal O}_\Delta$ having $\Delta > 0$. Construct the tensor product theory $\CNfold$ and quotient it by a permutation group $\ON \leq S_N$. There is a wide choice of such permutation groups; the class of permutation orbifold theories 
$\Cperm$ ranges from the  cyclic orbifold theory $\Ccyc$ to the symmetric orbifold theory  $\Csym$.  The latter is supposed to give a to a stringy holographic dual in the $N \gg 1$ limit, while the former interpolates into a  vector like classical higher spin theory. See Fig.~\ref{fig:plot} for a plot of the free energies where a sharp phase transition is clearly visible in case of the symmetric orbifold theory at large $N$, but not in the cyclic orbifold theory.

The question we pose is: for what choices of $\ON$ are we guaranteed to have a stringy holographic CFT in the $N\gg 1$ limit? This is really a group theoretic question, to answer which, we  formulate simple sufficiency condition building upon the result of \cite{Hartman:2014oaa}. The heuristic logic behind our analysis may be phrased as follows. In the $N$-fold tensor product one has $N$ times more states than in the seed theory ${\cal C}$. These need to be projected out to get a sparse low-lying spectrum. For simplicity consider the state obtained by acting with the lightest primary on $\ell$ out of the $N$ copies to get a state with dimension $\ell\,\Delta$ (we take $\ell \ll N$). The number of such states  depends on the number of orbits of $\ON$ on $\ell$-element subsets of ${\cal C}^N$; as long as it does not scale with $N$ for $\ell\ll N$ we have a sparse low lying spectrum. More specifically, we want the number of orbits of the permutation group $\ON$ on $\ell$-element subsets of an $N$-element set to remain finite as $N \to \infty$. In the strict limit, permutation groups exhibiting this property are called {\em oligomorphic permutation groups} \cite{Cameron:1990fk} (see also \cite{Cameron:2009rr} for some a more modern survey).\footnote{ The terminology is apposite: oligomorphy refers to the fact that the group has `few' orbits.} We will denote the strict $N\rightarrow \infty$ limit of a permutation group $\ON$ as $\Oinf$. 

While oligomorphic permutation groups allow for the existence of a sensible large $N$ limit, we still need to impose a further condition, to ensure that the spectrum is sparse enough. This turns out to be possible to do, once we bound the number of orbits of the group on (arbitrary) finite element subsets: we want the number of orbits on $\ell$ element subsets to grow at most exponentially in $\ell$ (with specific dependence set by the gap $\Delta$).  Without getting into too many technicalities at this point a succinct statement we may make is that   {\em Oligomorphic permutation orbifolds are (sometimes) holographic}.\footnote{ An oligomorphic permutation orbifold by itself leads to a sensible large $N$ limit, as the degeneracy of excited states remains bounded. However, should the low-lying degeneracy grow faster than an exponential then the Hawking-Page like phase transition is likely to be washed out. It would however be interesting to examine whether these orbifolds give rise to interesting models of the vector-type.}

The structure of this paper is as follows. In  \S\ref{sec:porbs} we review the construction of orbifold CFT partition functions and the group theory underlying it. The orbifold partition function is given by a sum over seed theory partition functions evaluated on unbranched covers of the torus (respecting the permutation action). We demonstrate this construction explicitly for cyclic and symmetric product theories, using them to build some intuition. In  \S\ref{sec:Torus} we extract a precise statement about  covering space geometries which dominate the partition function of orbifold CFTs in the limit of large degree permutation actions. Using these results, in \S\ref{sec:oligomorphs} we finally derive group theoretic conditions on the permutation group such that the orbifold has a sparse spectrum. A key tool we use in the process involves rewriting the orbifold partition sum in terms of the cycle index of the group in question.  Wreath product groups can be used to exemplify the construction of a large class of holographic CFTs which we describe in  \S\ref{sec:wreath}. We end end with a discussion in  \S\ref{sec:conclusion}. 

\medskip
\noindent
{\em Note added:} Reference \cite{Belin:2014fna} which appeared on the arxiv shortly after our work, discusses similar issues regarding permutation obrifold CFTs. We thank the authors for alerting us to their work. 

\section{A review of permutation orbifolds}
\label{sec:porbs}

Consider a CFT ${\cal C}$ characterized by its central charge $c \sim {\cal O}(1)$ and a discrete spectrum with a non-vanishing gap; the lowest primary associated with vertex operator ${\cal O}_\Delta$ having a conformal dimension $\Delta > 0$. The $N$-fold tensor product theory $\CNfold$ admits a natural action of permutation groups $\ON$ that act on $N$-element sets. Quotienting the tensor product theory by one such group $\ON$ results in a permutation orbifold theory
\begin{equation}
\Cperm \equiv \CNfold/\ON
\label{}
\end{equation}	
which will form the main focus of our investigation.  We will be interested in the $N \to \infty$ limit of these theories since this limit ensures that the central charge (which is independent of the choice of $\ON$)
\begin{equation}
\cc = N\, c \gg 1
\label{}
\end{equation}	
satisfy the first of our criteria.  The question we want to address is for what choice of $\ON$ is the second of our conditions, viz., sparseness of the low-lying spectrum satisfied.

It is useful to record some basic facts about permutation groups at this stage. Recall that the {\em degree} of a 
permutation group $\ON$ refers to the number of elements of the set the permutations act on. We will be interested in fixing the degree to $N \gg 1$ in what follows (and henceforth index our permutation group by its degree). On the other hand the {\em order} refers to the number of elements of the group (its cardinality as a set) and we will denote this by $|\ON|$. Intuitively, we may imagine that groups with more elements (larger order) lead to sparse spectra but the situation as we will see in a while is a bit more nuanced.

Since we are interested in the spectra of the orbifold theories, the simplest thing we can do is to examine the torus partition function. Let the partition function of the seed theory ${\cal C}$ be 
$Z(\tau,\btau)$ where $\tau = \tau_1 + i\, \tau_2$ is the modular parameter of torus. We will also use the parameterization 
\begin{align}
\tau_2 = \frac{\beta}{2\pi}\,, \qquad x = e^{-\beta}
\label{}
\end{align}	
when necessary to write simple expressions. The partition function of the orbifold theory $\Cperm$, denoted $Z_{N,\ON}(\tau,\btau)$, can be obtained from that of the seed theory $Z(\tau,\btau)$ by a nice group theoretic construction \cite{Bantay:1997ek}, which builds on the seminal work of \cite{Dijkgraaf:1996xw}. 

Before we explain the actual result, let us intuit physically what we should expect following the original construction of \cite{Dixon:1985jw}. Firstly, choose a canonical homology basis of $a$ and $b$ cycles for the torus on which we wish to compute $Z_{\perm}(\tau,\btau)$. We further recall that $\Cperm$ is obtained by gluing together $N$-copies of ${\cal C}$ with non-trivial elements of $\ON$ giving rise to twisted boundary conditions. Rather than considering $N$ fields related by (twisted) boundary conditions on a single torus, we can by a linear transformation pass to a basis where we consider a diagonal action of an element of $\ON$ on a twisted field. The twist is now simply prescribed by the monodromy picked up by the field as we take it around the $a$ or $b$ cycle of the torus. The twisted field naturally lives on the $N$-fold cover of the original torus. Furthermore, the covering space is unramified, i.e., there are no branch points for the monodromy action. This fact greatly simplifies the analysis of the torus partition function, since unramified covers of tori are again tori (by Riemann-Hurwitz). As a result, knowledge of $Z(\tau,\btau)$ is sufficient to determine $Z_{\perm}(\tau,\btau)$.

Let us now review the result of  \cite{Bantay:1997ek} which formalizes the above intuition. Consider homomorphisms from the fundamental group of the torus $\Gamma_1 = \fZ \oplus \fZ$ into the  group $\ON$,
\begin{align}
\phi: \Gamma_1 \to \; \ON \,.
\label{}
\end{align}
Of interest to us are the orbits of $\phi$ on the $N$-element set $X_N \equiv \{1,2,\ldots, N\}$ which can be denoted as 
\begin{align}
\mO(\phi) = \{ \phi(\Gamma_1)\cdot k \; | \;  k = 1,\ldots,N \} \,.
\label{}
\end{align}

The main result may now be stated as follows: consider all maps $\phi$, and for a given map focus on its orbits, i.e., the elements of $\mO(\phi)$. Each orbit can be associated with a new torus whose modular parameter $\tau_\xi$ is determined by the stabilizer subgroup of an element $\xi^*$ in the given orbit. This can be summarized in equations as\footnote{ In fact, the result stated here is quite general and can be applied directly to computing higher genus partition functions of permutation orbifolds  \cite{Bantay:1998fy}  as we review in Appendix \ref{sec:HigherGenus}.}
\begin{align}
Z_{\perm}(\tau,\btau) = \frac{1}{|\ON|} \, \sum_{\phi:\, \Gamma_1 \rightarrow \,\ON} \, 
\prod_{\xi \in \mO(\phi)} Z(\tau_\xi,\btau_\xi) \,,
 \label{eq:generalResultA}
\end{align}
where the modular parameters of the tori depend on the group theory data, viz.,
\begin{align}
\tau_\xi \equiv \tau[{S_\xi}]\,, \quad S_\xi \equiv \{ x\in \Gamma_1 \, | \, \phi(x) \xi^* = \xi^* \,\}  \text{ for any } \xi^* \in \xi \,.
\end{align}

Now for the torus a homomorphism $\phi: \Gamma_1 \rightarrow \ON$ is defined by its action on the generators $a$ and $b$ of $\Gamma_1$. Denote their 
images as $z_a = \phi(a)$, $z_b = \phi(b)$; any arbitrary assignment of this form gives a homomorphism as long as $z_a$ and $z_b$ are commuting elements in $\ON$. Therefore, 
\begin{align}
 Z_{\perm}(\tau,\btau) = \frac{1}{|\ON|} \sum_{\substack{z_a,z_b\in \ON \\ z_a z_b = z_b z_a}}  \ 
 \prod_{\xi \in \mO(z_a,z_b)} Z(\tau_\xi,\btau_\xi) \,.
 \label{eq:Z1general}
\end{align}

As described above, a homomorphism $\phi$ of the above type determines an unramified covering of the torus $\Sigma_1(\tau)$. In fact, the elements $z_a,\,z_b \in \ON$ determine how to move between 
sheets of the covering space as one moves around the $a$ and $b$-cycles of $\Sigma_1(\tau)$. If $\mO(z_a,z_b)$ contains a number of disjoint orbits, then the covering space consists of the same
number of connected components. Thus the product over $\xi \in \mO(z_a,z_b)$ is actually a product over the different connected components of the covering space. One such 
connected component covers again a torus but with modular parameter 
\begin{align}
\tau_\xi = \frac{\mu_\xi \, \tau + \kappa_\xi}{\lambda_\xi} \,,
\end{align}
where $\mu_\xi$ is the number of $z_a$ orbits contained in $\xi$, $\lambda_\xi$ is their common length, and $\kappa_\xi$ is the smallest non-negative integer for 
which $z_b^{\mu_\xi} = z_a^{\kappa_\xi}$.

To get a feeling for why we wish to focus on the sparse spectra, let us first construct and examine two simple (and very familiar) orbifolds; the cyclic and symmetric orbifolds,  for $\ON = {\mathbb Z}_N$ and $\ON = S_N$ respectively:
\begin{align}
\Ccyc \equiv\CNfold/{\mathbb Z}_N \,, \qquad \Csym\equiv \CNfold/S_N \,.
\label{}
\end{align}
Recall that $|{\mathbb Z}_N| = N$ and $|S_N| = N!$. Intuitively, this makes these groups the smallest and largest permutation groups with a transitive degree $N$ action. We will subsequently return to general $\ON$ for which we will need some further specification of the properties of the permutations involved.

\subsection{Cyclic  orbifolds}
\label{sec:CyclicFormalism}

For $\ON = \fZ_N$, the above considerations can easily be made explicit. We take $\fZ_N$ to be generated by a single element $z$: $\fZ_N = \{z^i\,, z^N = e\}$. The action of the group on an element $k$ of the $N$-set $X_N = \{1,2\,\ldots,N\}$ will be taken to be $z^i \cdot k = (k+i) \mod{N}$.

 The cyclic orbifold partition function  has been explored in \cite{Klemm:1990df,Borisov:1997nc}. For simplicity let us first consider $N$ to be a prime integer and record the basic result from these analysis
\begin{align}
Z_{\cyc}(\tau,\btau) &= \frac{1}{N} \left[ Z(\tau,\btau)^N + (N-1)\, \bigg\{ Z(N\,\tau, N\,\btau) 
+ \sum_{\kappa = 0}^{N-1} \; Z \left(\frac{\tau+ \kappa}{N},\frac{\btau+ \kappa}{N}\right)  \bigg\} \right] 
\notag \\
  &= \frac{1}{N} \left( T_1Z(\tau,\btau)\right)^N + (N-1) \, T_N Z_1(\tau,\btau) \,,  
  \label{eq:genus1Znresult}
\end{align}
where in the second line we have simplified the answer by introducing the Hecke operator.  For 
$k \in {\mathbb Z}_+$ the Hecke operator  maps modular forms into themselves. For our purposes the torus partition functions being modular invariant, the action of the $M^{\rm th}$-Hecke operator can be  defined in terms of the divisors of $M$: 
\begin{align}
  T_M Z(\tau,\btau) = \frac{1}{M} \sum_{d | M} \sum_{\kappa = 0}^{d-1}
  \ Z\left(\frac{M\,\tau + \kappa d}{d^2} ,
  \frac{M\,\btau + \kappa d}{d^2} \right) \,.
\end{align}

It is useful to see how this arises from the general formula given above in \eqref{eq:Z1general}. 
Consider homomorphisms $\phi: \Gamma_1 \equiv \fZ \oplus \fZ \rightarrow \fZ_N$ which are determined by their action on the generators $a$ and $b$ of $\Gamma_1$: $\phi(a) = z_a$, $\phi(b)=z_b$. Here $z_a$ and 
$z_b$ are arbitrary elements of $\fZ_N$ because the only condition (commutativity) is automatically fulfilled in 
$\fZ_N$. We can easily classify all possible choices of $z_a$ and $z_b$:
\begin{itemize}
 \item $z_a=z_b=e$: In this case an element of the joint orbit $\mO(z_a,z_b)$ leaves each element of $X_N$ fixed. As a result
any $\xi \in \mO(z_a,z_b)$ contains exactly one $z_a$ orbit and has $\mu_\xi = \lambda_\xi = 1$, $\kappa_\xi = 0$.  The corresponding contribution in the sum \eqref{eq:Z1general} is $Z(\tau,\btau)^N$, where the power of $N$ comes from $N$  possible choices of $\xi$. 
\item $z_a=e, \, z_b\neq e$: Now $\mO(z_a,z_b) = X_N$, but $z_a$ orbits are still of length $1$. Therefore the only $\xi \in\mO(z_a,z_b)$ 
  has $\mu_\xi = N$, $\lambda_\xi = 1$, $\kappa_\xi = 0$. Accounting for $N-1$ different choices for $z_b$, we get a contribution to (\ref{eq:Z1general}) of   the form $(N-1)\, Z(N\tau, N\btau)$.
\item $z_a\neq e, \, z_b=e$: In this case $\mO(z_a,z_b) = X_N$ with $z_a$ orbits now having length $N$, i.e., 
  $\mu_\xi = 1$, $\lambda_\xi = N$, $\kappa_\xi = 0$. The contribution to the partition function is 
  $(N-1) \, Z_1(\tfrac{\tau}{N} , \tfrac{\btau}{N})$.
\item $z_a, z_b \neq e$: Again we have $\mO(z_a,z_b) = X_N$ with the $z_a$ orbits still being of length $1$. However, now $\kappa_\xi$ now runs from $1$ to $N-1$, depending on the choice of $z_b \in \{z_a, z_a^2, \ldots, z_a^{N-1}\}$, where we used that $\fZ_N$ for $N$ prime can be written as generated by $z_a$). The contribution to the partition function is therefore
  \begin{align}
    (N-1) \sum_{\kappa = 1}^{N-1} Z \left(\frac{\tau+ \kappa}{N}, \frac{\btau+ \kappa}{N}\right) \,,
  \end{align}
  where the factor $(N-1)$ comes from the different but equivalent choices for $z_a$.\footnote{ Note that we can include into this sum the contribution $(N-1) \, Z(\tfrac{\tau}{N}, \tfrac{\btau}{N})$ corresponding to the choice $z_b =e$ by letting the summation index $\kappa$ start from $0$ instead.}
\end{itemize}
If we sum up the enumerated contributions, the expression \eqref{eq:Z1general} reduces to the answer quoted above in \eqref{eq:genus1Znresult}.

The generalization of the above discussion to non-prime $N$ looks as follows:\footnote{ This can be simplified a bit in terms of the Euler totient function; this is also more natural from the cycle index of ${\mathbb Z}_N$ which we will have more to say about in \S\ref{sec:oligomorphs}.}
\begin{align}
  Z_{\cyc}(\tau) &= \frac{1}{N} Z(\tau)^N + \frac{1}{N}\sum_{r=1}^{N-1} \left[ Z \left(\frac{(N,r)}{N} \tau \right)^{(N,r)} + Z\left( \frac{N}{(N,r)} \tau\right)^{(N,r)} \right] \notag\\
      &\quad +\frac{1}{N} \sum_{r=1}^{N-1} \sum_{s=1}^{N-1} Z \left( \frac{(N,r)}{N} \left( \frac{(N,r)}{(N,r,s)} \tau + \kappa(r,s) \right)\right)^{(N,r,s)} \notag\\
      &= \frac{1}{N} \sum_{r=1}^{N} \sum_{s=1}^{N} Z \left( \frac{(N,r)}{N} \left( \frac{(N,r)}{(N,r,s)} \tau + \kappa(r,s) \right)\right)^{(N,r,s)} \,,
    \label{eq:Z1NonPrime}
\end{align}
where $(p,\ldots,q) \equiv \text{gcd}(p,\ldots,q)$ and $\kappa(r,s)$ is defined as the \textit{smallest} integer in $\{0,1,\ldots,\tfrac{N}{(N,r)}-1\}$ such that
$\left(\kappa(r,s)\, r - \frac{(N,r)\, s}{(N,r,s)}\right) = 0\, \text{mod} \, N $.
The first two lines in (\ref{eq:Z1NonPrime}) resemble the structure of the result (\ref{eq:genus1Znresult}) and one can easily check that for $N$ prime the two formulae are the same (the minimization parameter in the definition of $\kappa(r,s)$ vanishes for prime $N$). The general result as far as we are aware doesn't admit a nice rewriting in terms of Hecke operators.

\paragraph{Example:}
As the simplest non-trivial illustration of the result (\ref{eq:Z1NonPrime}), consider the case $N=4$. 
Table \ref{table:Z4} lists all terms that appear in the 
partition function $Z_{4,\fZ}$ and the data $(\mu,\lambda,\kappa)$ which give their corresponding modular parameters. 
For instance, the bottom right corner of the table corresponds to the term $Z_1(\tau)^4$, i.e., the case where both $x$ and $y$ are trivial.
The resulting partition function reads
\begin{align}
  Z_{4,\fZ} &= \frac{1}{4} \bigg[ Z_1\left(\tau\right)^4 + Z_1\left(2\tau\right)^2 + Z_1\left(\frac{\tau}{2}\right)^2 
      + Z_1\left(\frac{\tau+1}{2}\right)^2 + 2 Z_1\left(\frac{2\tau + 1}{2} \right)\notag \\
  &\quad + 2 Z_1\left(4\tau\right) + 2 Z_1\left(\frac{\tau}{4}\right) + 2 Z_1\left(\frac{\tau+1}{4} \right) + 2 Z_1\left(\frac{\tau+2}{4} \right) 
         + 2 Z_1\left(\frac{\tau+ 3}{4}\right)
            \bigg] \,. 
  \label{eq:Z4covers}
\end{align}

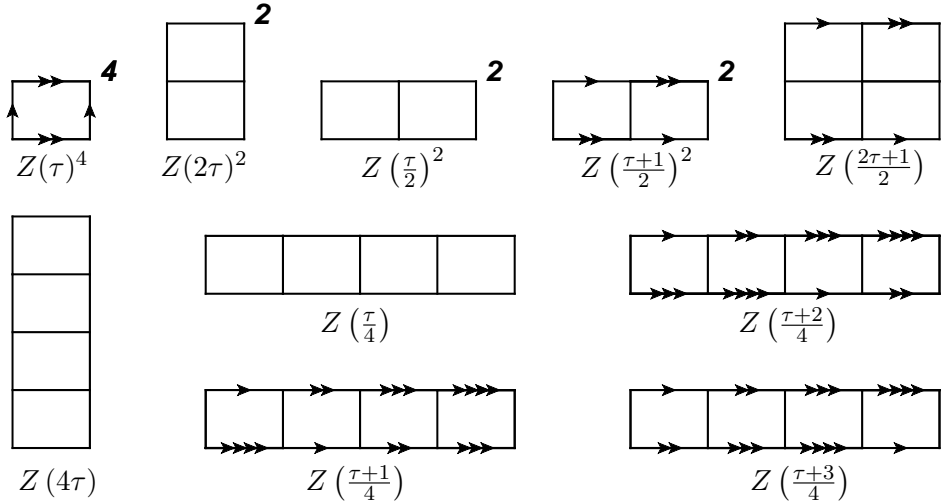
\begin{figure}
 \centerline{\includegraphics[width=.8\textwidth]{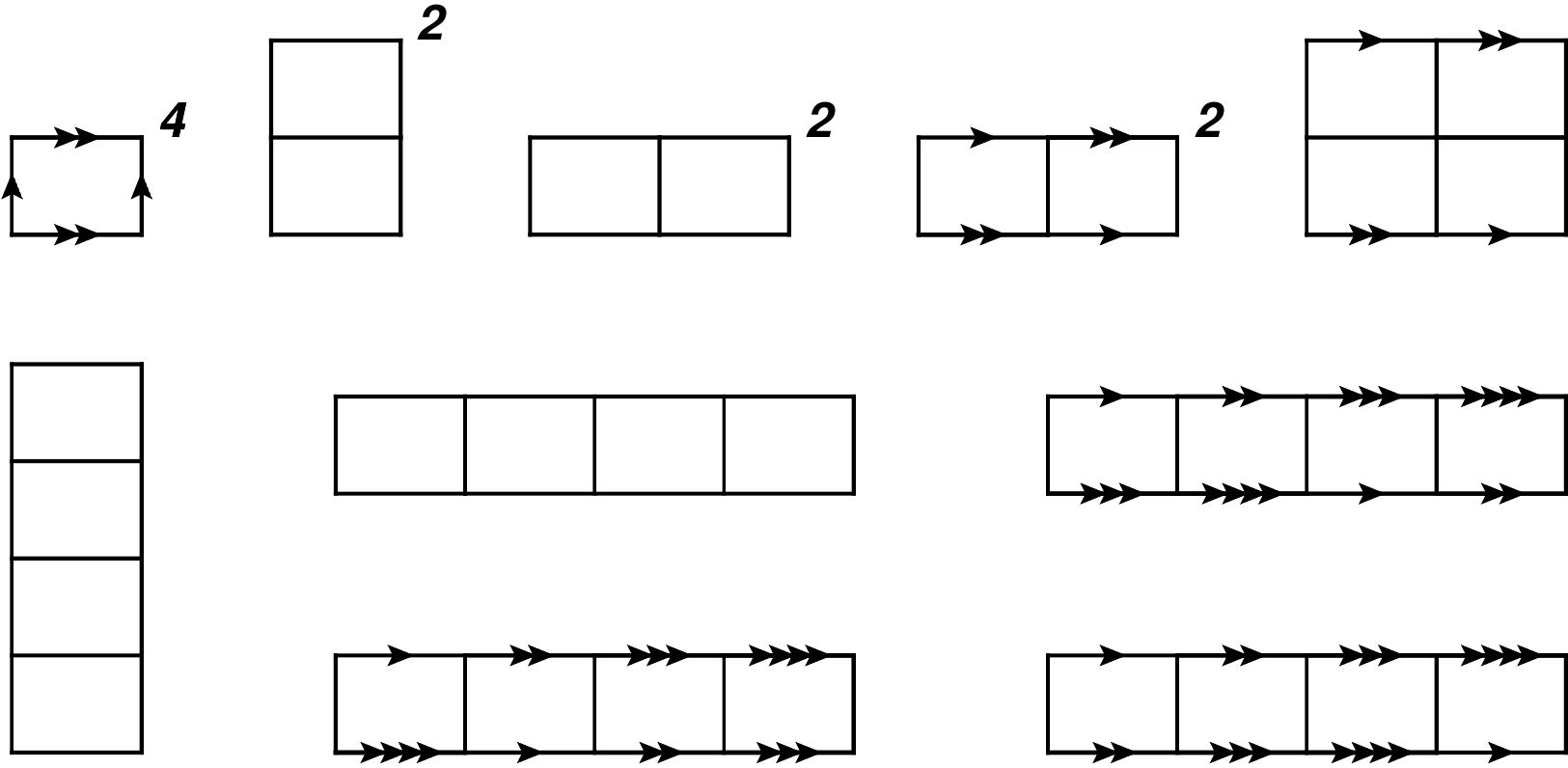}}
  \setlength{\unitlength}{0.1\columnwidth}
 \begin{picture}(0.3,0.4)(0,0)
 \put(1.4,2.9){\makebox(0,0){$Z(\tau)^4$}}
 \put(2.7,2.9){\makebox(0,0){$Z(2\tau)^2$}}
 \put(4.4,2.9){\makebox(0,0){$Z\left(\frac{\tau}{2}\right)^2$}}
 \put(6.4,2.9){\makebox(0,0){$Z\left(\frac{\tau+1}{2}\right)^2$}}
 \put(8.4,2.9){\makebox(0,0){$Z\left(\frac{2\tau+1}{2}\right)$}}
 \put(1.45,.2){\makebox(0,0){$Z\left(4\tau\right)$}}
 \put(4,.2){\makebox(0,0){$Z\left(\frac{\tau+1}{4}\right)$}}
 \put(7.8,.2){\makebox(0,0){$Z\left(\frac{\tau+3}{4}\right)$}}
 \put(4,1.55){\makebox(0,0){$Z\left(\frac{\tau}{4}\right)$}}
 \put(7.7,1.55){\makebox(0,0){$Z\left(\frac{\tau+2}{4}\right)$}}
 \end{picture}
 \caption{Illustration of the content of (\ref{eq:Z4covers}). Every graph consists of 4 boxes each representing one sheet of a $4$-fold cover of the torus with modular parameter $\tau$. Depending on how the sheets are sewn together, we obtain different covering spaces which are all tori with different modular parameters (edges without arrows are glued to the opposite edge). The partition function $Z(\tau,\btau)$ of a given parent CFT ${\cal C}$ has to be evaluated on all these covers in order to get the partition function of the orbifold theory ${\cal C}^{\otimes 4}/\fZ_4$. This can be seen from the fact that each of the covers has an automorphism group of sheet permutations that is generated by a set of elements of $\fZ_4$.  
 }
 \label{fig:Z4covers}
 \end{figure}

Note that the first term by itself is obviously modular invariant. Furthermore, the following three terms together are modular invariant; and thus all the remaining terms taken together are modular invariant. An illustration of the various contributions can be found in Fig.~\ref{fig:Z4covers}. This simply illustrates the earlier observation that the relevant geometries are just all unbranched $4$-sheeted covers of the torus that have an automorphism group of sheet permutations which is generated by a set of $\fZ_4$ elements. 
\begin{table}[t]
 \small
 \centering
{\tabulinesep=1.5mm
\begin{tabu}{|c||c|c|c|c|}
 \hline
     & $s=1$ & $s=2$ & $s=3$ & $s=4$
      \\ 
 \hline\hline 
 \multirow{2}{1cm}{$r=1$} & $\mO=\{\{1,2,3,4\}\}$& $\mO=\{\{1,2,3,4\}\}$ & $\mO=\{\{1,2,3,4\}\}$ & $\mO=\{\{1,2,3,4\}\}$ \\
                          & $(\mu,\lambda,\kappa)=(1,4,1)$ & $(\mu,\lambda,\kappa)=(1,4,2)$ & 
                            $(\mu,\lambda,\kappa)=(1,4,3)$ & $(\mu,\lambda,\kappa)=(1,4,0)$ \\
 \hline
 \multirow{2}{1cm}{$r=2$} & $\mO=\{\{1,2,3,4\}\}$& $\mO=\{\{1,3\},\{2,4\}\}$ & $\mO=\{\{1,2,3,4\}\}$ & $\mO=\{\{1,3\},\{2,4\}\}$ \\
                          & $(\mu,\lambda,\kappa)=(2,2,1)$ & $(\mu,\lambda,\kappa)=(1,2,1)$ & 
                            $(\mu,\lambda,\kappa)=(2,2,1)$ & $(\mu,\lambda,\kappa)=(1,2,0)$ \\
 \hline
 \multirow{2}{1cm}{$r=3$} & $\mO=\{\{1,2,3,4\}\}$& $\mO=\{\{1,2,3,4\}\}$ & $\mO=\{\{1,2,3,4\}\}$ & $\mO=\{\{1,2,3,4\}\}$ \\
                          & $(\mu,\lambda,\kappa)=(1,4,3)$ & $(\mu,\lambda,\kappa)=(1,4,2)$ & 
                            $(\mu,\lambda,\kappa)=(1,4,1)$ & $(\mu,\lambda,\kappa)=(1,4,0)$ \\
 \hline
 \multirow{2}{1cm}{$r=4$} & $\mO=\{\{1,2,3,4\}\}$& $\mO=\{\{1,3\},\{2,4\}\}$ & $\mO=\{\{1,2,3,4\}\}$ & $\mO=\{\{1\},\{2\},\{3\},\{4\}\}$ \\
                          & $(\mu,\lambda,\kappa)=(4,1,0)$ & $(\mu,\lambda,\kappa)=(2,1,0)$ & 
                            $(\mu,\lambda,\kappa)=(4,1,0)$ & $(\mu,\lambda,\kappa)=(1,1,0)$ \\
 \hline
\end{tabu}}
 \caption{Combinatorics for the computation of $Z_{4,\fZ}(\tau,\btau)$. 
 The integers $r,s$ are elements of 
 $\fZ_4$ which we now take to be the additive group with elements $\{1,2,3,4\}$ and an obvious action on an integer $k$ via $r\cdot k = (k+r)\, \text{mod} \, 4$. The set of orbits of $r$ and $s$ is denoted by $\mO$ and the torus modular parameter in each case is given by $\frac{\mu \tau + \kappa}{\lambda}$.}
 \label{table:Z4}
 \end{table}

As an aside, let us note that cyclic orbifolds can easily be treated in this formalism at higher genus, as well. The logic is very similar to what we have demonstrated for the torus: the orbifold partition function is given by a sum of products of the parent CFT's partition function evaluated on unbranched covers of the given Riemann surface. However, since unbranched covers of a genus $g$ Riemann surface can have genus higher than $g$, some qualitatively new complications appear. We review the basic formalism of how this works in Appendix \ref{sec:CycHigherGenus} and postpone a more detailed analysis for later \cite{Haehl:2015gf}.

\subsection{Symmetric orbifolds}
\label{sec:symorb}

For the case of $\ON$ being the full symmetric group $S_N$, the orbifold partition function was derived 
originally in a beautiful analysis by \cite{Dijkgraaf:1996xw}. In this case it is actually easier to give a generating function for the $S_N$ orbifold (for the same reason that it is simpler to present the grand canonical partition function for particles obeying Bose statistics). This is given succinctly in terms of Hecke operators as
\begin{align}
\sum_{N=0}^\infty \, t^N\, Z_{\sym}(\tau,\btau) = \exp\left(\sum_{M=1}^{\infty} \, t^M\, T_M Z(\tau,\btau) \right)
\label{eq:symgent}
\end{align}
where $t$  (the fugacity) is an auxiliary variable introduced to write the generating function.

This expression may also be obtained from the general result quoted in \eqref{eq:Z1general} using the following logic.  First we use the fact that the sum over connected covers of the torus can be equivalently understood in terms of a sum over the finite index subgroups of the fundamental group \cite{Bantay:2000eq}. On the torus, the sum over finite index subgroups is just the Hecke operator, i.e., we can write\footnote{ This argument also extends straightforwardly to higher genus orbifolds (see Appendix \ref{sec:HigherGenusSN}).}
\begin{align}
 Z_{\sym}(\tau,\btau) = \frac{1}{N!} \sum_{z\in S_N} \;\prod_{\xi \in \mO(z)} |\xi| \, T_{|\xi|}(\tau,\btau) \,.
 \label{eq:ZSn}
\end{align}
Now observe that only the length of $\xi$ matters in this expression. Therefore the relevant information contained in $z\in S_N$ is the number of cycles of a given length. Denote a generic element $z\in S_N$ which contains $m_k$ cycles of length $k$ by 
\begin{align}
  z = (1)^{m_1} (2)^{m_2} \cdots (N)^{m_N} \,, \qquad \sum_{k=1}^N k \, m_k = N \,.
\end{align}
We shall refer to such an element $z$ as being of \textit{cycle type} $\{m_k\}_N \equiv\{m_1,\ldots, m_N\}$. The number of different elements in $S_N$ which are of
cycle type $\{m_k\}_N$ is 
\begin{align}
  d_{\{m_k\}_N} = N! \prod_{k=1}^N  \frac{ 1}{k^{m_k}\, m_k!} \,. 
\end{align}
We can therefore write the genus one symmetric orbifold partition function \eqref{eq:ZSn} as\footnote{ Once again this rewriting is a lot more natural in terms of the cycle index of $S_N$, cf., \S\ref{sec:oligomorphs}.} 
\begin{align}
 Z_{\sym}(\tau,\btau) 
   &=  \frac{1}{N!} \,\sum_{\{m_k\}_N} d_{\{m_k\}_N}\, \prod_{k=1}^N 
   \left[ k \, T_kZ(\tau,\btau)\right]^{m_k} \notag \\
   &= \sum_{\{m_k\}_N} \prod_{k=1}^N \; \frac{ 1}{ m_k!}\, \left[T_kZ(\tau,\btau)\right]^{m_k} \,,
  \label{eq:GeneralSnresult}
\end{align}
where the sum runs over all possible cycle types, i.e., over all sets of integers $\{m_1,\ldots,m_N\}$ which satisfy 
$\sum_k \, k \,m_k = N$. 

It is rather immediate to check from here that the expression for the generating function obtained by multiplying the two sides by $t^N$ and summing over $N \in {\mathbb Z}_+$ leads to the beautiful expression \eqref{eq:symgent}. 

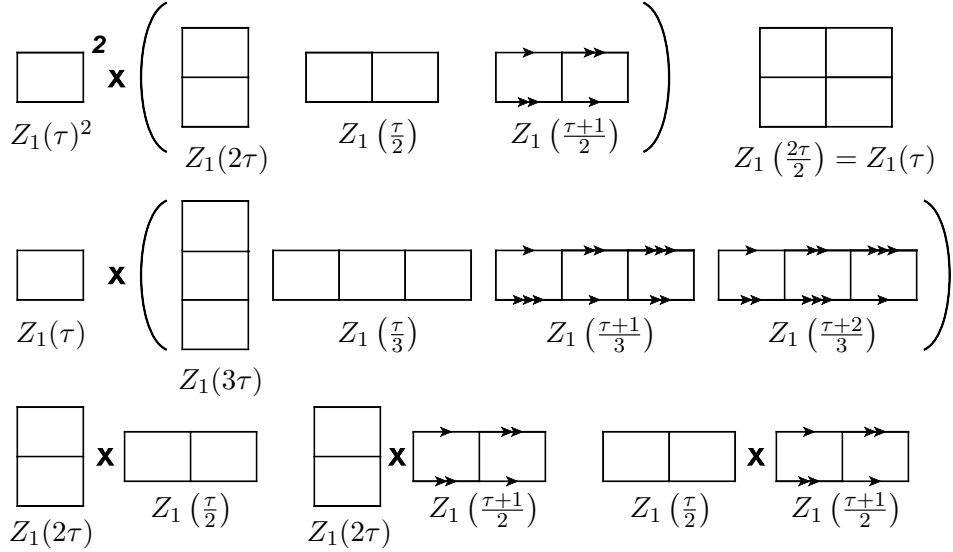
\begin{figure}
 \centerline{\includegraphics[width=.8\textwidth]{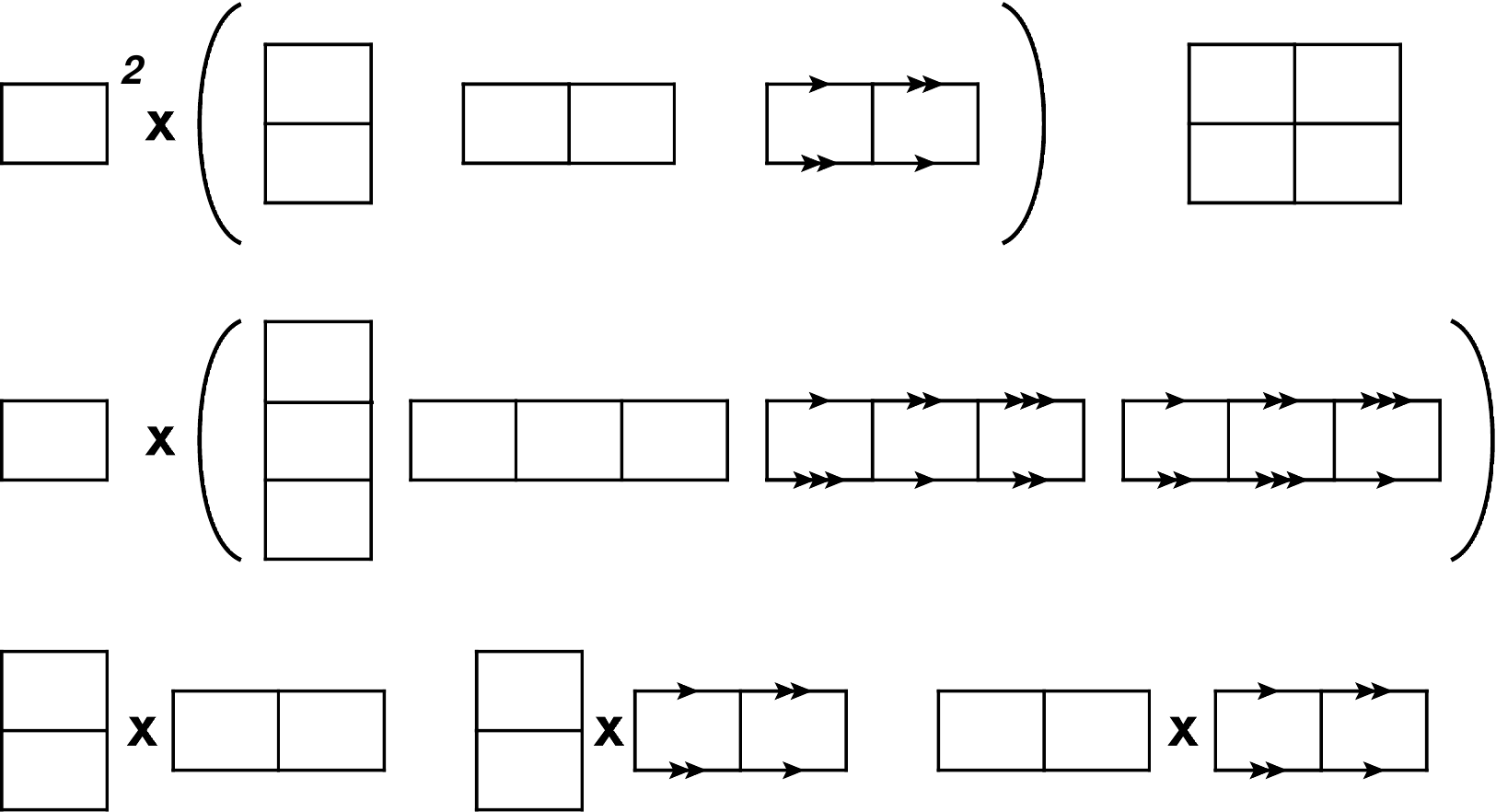}}
  \setlength{\unitlength}{0.1\columnwidth}
 \begin{picture}(0.3,0.4)(0,0)
 \put(1.3,3.6){\makebox(0,0){$Z_1(\tau)^2$}}
 \put(2.8,3.4){\makebox(0,0){$Z_1(2\tau)$}}
 \put(4.1,3.6){\makebox(0,0){$Z_1\left(\frac{\tau}{2}\right)$}}
 \put(5.7,3.6){\makebox(0,0){$Z_1\left(\frac{\tau+1}{2}\right)$}}
 \put(8,3.4){\makebox(0,0){$Z_1\left(\frac{2\tau}{2}\right) = Z_1(\tau)$}}
 \put(1.3,1.9){\makebox(0,0){$Z_1(\tau)$}}
 \put(2.75,1.5){\makebox(0,0){$Z_1(3\tau)$}}
 \put(4.1,1.9){\makebox(0,0){$Z_1\left(\frac{\tau}{3}\right)$}}
 \put(6,1.9){\makebox(0,0){$Z_1\left(\frac{\tau+1}{3}\right)$}}
 \put(7.9,1.9){\makebox(0,0){$Z_1\left(\frac{\tau+2}{3}\right)$}}
 \put(1.3,.2){\makebox(0,0){$Z_1(2\tau)$}}
 \put(2.5,.4){\makebox(0,0){$Z_1\left(\frac{\tau}{2}\right)$}}
 \put(3.85,.2){\makebox(0,0){$Z_1(2\tau)$}}
 \put(5,.4){\makebox(0,0){$Z_1\left(\frac{\tau+1}{2}\right)$}}
 \put(6.6,.4){\makebox(0,0){$Z_1\left(\frac{\tau}{2}\right)$}}
 \put(8.1,.4){\makebox(0,0){$Z_1\left(\frac{\tau+1}{2}\right)$}}
 \end{picture}
 \caption{Box diagrams of all eleven $4$-sheeted unbranched covers of a torus whose automorphism groups of sheet permutations are not generated by elements of $\fZ_4$. Together with the geometries of Fig.~\ref{fig:Z4covers} these form the complete set of all $4$-sheeted unbranched covers of the torus. All of the these 
 $21$ covers  are relevant for computing the symmetric orbifold partition function. }
 \label{fig:S4covers}
 \end{figure}

\paragraph{Example:} 
As an example consider again $N=4$. In  \S\ref{sec:CyclicFormalism} we gave an explicit construction of the cyclic product orbifold partition function $Z_{4,{\mathbb Z}}(\tau,\btau)$. We saw that the latter is essentially given by the sum over products of $Z(\tau,\btau)$ evaluated on $4$-sheeted unbranched covers of the torus which are, of course, again tori but at different points in moduli space. In Fig.~\ref{fig:Z4covers} we demonstrated that all these covers had an automorphism group of sheet permutations that are generated by elements of $\fZ_4$. 
We are now however interested in the $S_4$ orbifold theory. Thus we  would naturally expect that its partition function is given in terms of  (products of) seed partition functions evaluated on all unbranched covers of the torus which have an automorphism group that is consistent with $S_4$ instead. Indeed, directly applying \eqref{eq:GeneralSnresult} for $N=4$, yields a somewhat long expression which, however, contains exactly the expected products of torus partition functions. The relevant geometries now exhaust the set of all unbranched $4$-sheeted covers of the torus, 
i.e., both cyclic ones as in Fig.~\ref{fig:Z4covers} and non-cyclic ones which we list in Fig.~\ref{fig:S4covers} for completeness. 
 %

\section{Spectral properties of cyclic and symmetric orbifolds}
\label{sec:Torus}

We  described in \S\ref{sec:porbs} the general formalism for dealing with permutation orbifolds. More specifically we  explained how to obtain the partition functions for the cyclic and symmetric groups for some fixed degree $N$. We now turn to examining the behaviour as a function of $N$. Of specific concern to us will be the asymptotics in the  $N \to \infty$ limit when we attain large central charge 
($\cc \equiv Nc \to \infty$). 

\subsection{Example: The free boson orbifold}
\label{sec:freeb}

To gain some intuition let us first look at a simple example. Consider as our seed theory, the $c=1$ free boson compactified on a (spatial) circle of radius $R=1$ (where it is dual to a Dirac fermion). The partition function of this theory is well known and is given as
\begin{align}
  Z^b(\tau,\bar \tau) &= 
  (q \bq)^{-\frac{1}{24}} |\eta_2(\tau)| \sum_{e,m\in \fZ} q^{\frac{1}{2} \left(e + \frac{m}{2}\right)^2}
     \bq^{\frac{1}{2} \left(e - \frac{m}{2}\right)^2} \,,
 \label{eq:bosonR1}
\end{align}
where we have an unconventional definition of the infinite product 
\begin{equation}
\eta_M(q) =  \prod_{k=1}^\infty \; \frac{1}{ (1-q^k)^M}  \,.
\label{eq:etaM}
\end{equation}	
It will be useful for our purposes to focus on a rectangular torus with $\tau = i\,\tfrac{\beta}{2\pi}$ where $\beta = T^{-1}$ is the inverse temperature.

Using this seed partition function it is easy to numerically examine the behaviour of the rescaled vacuum subtracted free energy
\begin{align}
\Delta F(T) = \frac{1}{N}\ \left(-T\, \log Z(\beta) + \frac{\cc}{12} \right)
\label{eq:deltaF}
\end{align}
in the asymptotic limit. The result of this exercise is shown in Fig.~\ref{fig:plot} for the cyclic and symmetric orbifolds respectively.  From this plot it is rather easy to infer that 
\begin{equation}
\Delta F_{\sym}(T) =
\begin{cases}
& {\cal O}(N^{-1}) \,,\qquad\quad\, T < \frac{1}{2\pi} \\
& {\cal O}(1) \, T^2 \,,\qquad \quad\; T \geq \frac{1}{2\pi}
\end{cases}
\label{}
\end{equation}	
for the symmetric orbifold indicating a sharp large $N$ phase transition at $T=\frac{1}{2\pi}$ as expected from earlier analysis \cite{Keller:2011xi}. Indeed this is the behaviour we expect to see from two-derivative gravity theories in an asymptotically \AdS{3} spacetime with the low temperature phase being governed by the thermal AdS geometry and the high temperature phase by the BTZ black hole. Of particular note in this case is the fact that the vacuum contribution dominates all the way to $T =\tfrac{1}{2\pi}$ illustrating the sparseness of the spectrum. The high temperature $T^2$ growth is of course understood to be a consequence of Cardy scaling (due to modular invariance).

 On the other hand the cyclic orbifold exhibits a free energy which shows no sharp feature as $N \gg 1$ and smoothly interpolates between the low temperature vacuum dominated phase to the high temperature phase where $\Delta F(T) \sim T^2$. This is in accord with our earlier intuition that these theories do not give rise to local gravitational duals \cite{ElShowk:2011ag}.

It is in fact illustrative to examine the partition function itself as a function of temperature in some detail. Expanding out \eqref{eq:bosonR1} in Fourier series in the variable $x=e^{-\beta}$ we have
\begin{align}
Z^b(x) = x^{-\frac{1}{12}}\, \left(1 +2\, x^{\frac{1}{4} } + 6\, x + \cdots \right) .
\label{eq:ZbR1}
\end{align}
One finds by straightforward computation the symmetric orbifold partition function
\begin{align}
 Z^b_{\sym}(x) = x^{-\frac{N}{12}} \big[ 1 &+ x^{\frac{1}{8}}+x^{\frac{2}{9}} + 5 x^{\frac{1}{4}} + 2 x^{\frac{11}{36}}  
    + x^{\frac{5}{16}} + x^{\frac{25}{72}} + 7 x^{\frac{3}{8}} + x^{\frac{2}{5}} \notag \\
    & + 2 x^{\frac{31}{72}} + x^{\frac{7}{16}} + x^{\frac{4}{9}} + 2 x^{\frac{9}{20}} + 5 x^{\frac{17}{36}}+
    + x^{\frac{35}{72}} + 17 x^{\frac{1}{2}}+ \mO(x^{\frac{1}{2}+\eps}) \big] \,.
    \label{eq:fbsym}
\end{align}
These are all terms up to $\mO(x^{1/2})$ and their coefficients are independent of $N$
for all $N\geq 8$.\footnote{ The states associated with operator dimensions $\frac{1}{8}, \frac{2}{9}$, etc., arise from the twisted sectors of the orbifold.} For smaller values of $N$, we see some mild dependence in the coefficients on $N$, but as $N$ grows states with increasingly high conformal weight freeze out and therefore become insignificant as $N\rightarrow \infty$. This behaviour of the density of states is one of the characteristic features of the symmetric orbifold and is responsible for the sparseness of the low-lying spectrum. 

On the other hand there is no such freeze-out in the cyclic orbifold. While it is not trivial to write down a universal result (owing to the number theoretic dependence on the degeneracy of states), it is once again instructive to examine the behaviour of the partition function for prime $N$. In this case it is easy to show that the cyclic orbifold partition function behaves as
\begin{align}
 Z^{b}_{\cyc}(x) &= x^{-\frac{N}{12}} \left[ 1 + 2\, x^{\frac{1}{4}} + 2\,(N-1) \,x^{\frac{1}{2}} + \ldots \right] \,,
 \label{eq:fbcyc}
\end{align}
The linear growth of the degeneracy of the state with dimension $\frac{1}{2}$ with $N$ is in fact responsible for the non-sparseness in the spectrum. It is easy to argue that this behaviour arises from acting with two copies of the lightest primary vertex operator  (each with weight $\frac{1}{4}$) on two distinct copies of ${\cal C}$ in the $N$-fold tensor product. Given the two-fold degeneracy of the operator from \eqref{eq:ZbR1}, counting the number of ways to act with it on two distinct copies consistent with $\fZ_N$, one arrives at the correct degeneracy. This can equivalently be attributed to the fact that there are $\frac{1}{2}(N-1)$ orbits of $\fZ_N$ on two-element subsets of $X_N$, and $(N-1)$ orbits on ordered two element subsets of $X_N$, 
which we will make extensive use of later.

\subsection{Asymptotic behaviour of cyclic and symmetric orbifolds}
\label{sec:asycs}

While it is instructive to have some intuition from explicit analysis of a simple seed theory, it is useful for the generalizations we have in mind to consider abstracting this result. Let us therefore extract the essential features of the two cases. We find it convenient to break up the discussion into two distinct parts: (a) estimating the dominant contributions at low and high temperatures and (b) delineating the contribution from excited states.

The salient features of the partition function can be encapsulated in three basic results which we can summarize in the following three theorems. We start with a claim about the geometries relevant for the computation of the high and low temperature behavior of torus partition function. Subsequently we consider the large $N$ asymptotics at intermediate temperatures for the two orbifolds of interest separately.  

\begin{theorem} \label{thm:0}
  At large large $N$, the geometries which are dominant in the torus partition function of $\fZ_N$ and $S_N$ orbifold CFTs  are obtained as follows. Take $N$ numbered copies of the torus with modular parameter 
  $\tau$ and arrange them in any order in any number of groups which is consistent with cycle structure of the symmetry group $\fZ_N$ or $S_N$, respectively. Within each group either leave all tori disconnected, or sew together {\it all} of them, either along a-cycles or along b-cycles, without twisting the sewn tori with respect to each other.
\end{theorem}

\begin{theorem} \label{thm:1}
 Given a conformal dimension $\Delta_{max}$ there exists $N_*(\Delta_{max},c)$ such that the partition function $Z_{\cyc}(x)$ is universal for $N > N_*$ up to the order $\mO\left(x^{\Delta_{max}}\right)$. 
The  stable form for the partition function is given by
\begin{align}
  Z_{\cyc}(x) = \frac{1}{N} \left[ Z(x)^N +  
   \sum_{r=1}^{N-1} Z\left(\frac{N}{(N,r)}\,x\right)^{(N,r)}\right] +  x^{-\frac{\cc}{12}}\; \mO\left( x^{\Delta_{max}}\right)\,, 
 \label{eq:ThmZ}
 \end{align}
 with the value of $N_*$ determined as
\begin{align}
 N_* &= \frac{16\, \Delta_{max}}{c}
 \label{eq:Nstar}
\end{align}
 If we restrict attention to prime $N$ then we have some simplifications for we only get contributions from untwisted sector states
\begin{align}
 Z_{\cyc}(x) = \frac{1}{N} \bigg[ Z(x)^N + (N-1) \, Z(N \,x)\bigg]+ x^{-\frac{\cc}{12}} \;\mO\left( x^{\Delta_{max}}\right)\,, 
 \end{align}
though we pay a price as $N_*$ depends in more detail on the spectrum of the seed theory:
\begin{align}
N_* =\max\left\{\frac{\Delta_{max}}{h_{min}}\,, \,   \frac{6 \Delta_{max}}{c}+\sqrt{\left(\frac{6 \Delta_{max}}{c}\right)^2 + 1} \right\} \,,
\label{eq:ThmZ2}
\end{align}
where $h_{min}$ is the minimum eigenvalue of either $L_0$ or ${\bar L}_0$, i.e., $h_{min} = \min\{h, {\bar h}\}$.
\end{theorem}

\begin{theorem}  \label{thm:2}
 Given a conformal dimension $\Delta_{max}$ there exists $N_*$ such that the (rescaled) partition function $Z_{\sym}(x)$ is universal for $N > N_*$ up to the order $\mO\left(x^{\Delta_{max}}\right)$, i.e.,
 \begin{align}
   x^{\frac{Nc}{12}} \, Z_{\sym}(x) - x^{\frac{(N-1)c}{12}} \, Z_{(N-1),S}(x) =
   \mO\left(x^{\Delta_{max}}\right) \,,
   \label{eq:thmS}
 \end{align}
with 
\begin{align}
N_* = \max\left\{\frac{16}{c}\Delta_{max},\,\frac{\Delta_{max}}{h_{min}}\right\}
\label{}
\end{align}
\end{theorem}

The proofs of these statements are straightforward once we estimate the contribution of the individual Hecke operators. We relegate the technicalities to Appendix \ref{sec:asymproofs} and focus here on some general lessons we can learn.

To get a feeling for the above results,  consider truncating the partition function $Z(x)$ at the first non-trivial state with smallest dimension $\Delta_1$, i.e.,
\begin{align}
  Z(\tau,\bar \tau) = (q\bq)^{-\frac{c}{24}} \left[ 1 + \dee q^{\frac{1}{2}\Delta_1} \bq^{\frac{1}{2}\Delta_1} + \ldots\right] .
  \label{eq:Z1truncated}
\end{align}
Here $d_1\in {\mathbb Z}_+$ is the degeneracy of the state in the seed theory.

Using the result of  Theorem \ref{thm:1} we then learn that for  sufficiently large (prime) $N$ the cyclic orbifold partition function reduces to
\begin{align}
Z_{\cyc}(x) = x^{-\frac{cN}{12}} \left[ 1 + \dee \, x^{\Delta_1} + \frac{(N-1)}{2!} \dee^2 \, x^{2 \Delta_1}
   + \frac{(N-1)(N-2)}{3!} \dee^3 \, x^{3\Delta_1} + \ldots \right] .
   \label{eq:Z1LowestExpand}
\end{align}
We see here that the low-lying states whose dimension is independent of  $N$ have a degeneracy which grows polynomially in $N$. This is the reason why the cyclic orbifold partition function has no sharp phase transition. In fact, the degeneracies are related to the number of orbits of $\fZ_N$ on $k\leq N$-element subsets of $X_N$; we will rephrase the orbifold computation to make this manifest below.
Note that \eqref{eq:Z1truncated} only captures the contributions from $x^{\Delta_1}$. Higher weight states may interfere with this result if we have states with energies below $k\,\Delta_1$ for some $k$.

We can compare this to the low-lying spectrum of the $S_N$-orbifold theory. Theorem \ref{thm:2} guarantees
that for a given CFT the low-lying sector of $Z_{\sym}(x)$ is independent of $N\gg1$. However, we should bear in mind that there are still low-lying states; only their degeneracy freezes out and is fixed beyond a certain $N$ (leading to their insignificance asymptotically at large $N$). Another way of saying this is that all but a finite number of low-lying states get projected out from 
the tensor product theory $\mC^{\otimes N}$ after orbifolding. A closed form analogous to (\ref{eq:Z1LowestExpand}) cannot easily be given because the coefficients depend very much on the underlying CFT.\footnote{ The situation is well exemplified by our  result for the free boson \eqref{eq:fbsym} where we have the terms  up to $\mO(x^{1/2})$ -- the degeneracies are  independent of $N$ for all $N\geq N_*=8$ as predicted.}
Physically,  given a cutoff order $\Delta_{max}$, one only needs to keep track of a finite number of (well-defined) twisted sectors of Hecke sums $T_kZ(x)$ with small $k$. For $k$ sufficiently large, one may truncate the Hecke sum at its leading untwisted sector contribution, cf., \eqref{eq:HeckeTruncate}. In this sense the symmetric orbifold theories are essentially only sensitive to
untwisted sectors in the limit $N\rightarrow \infty$. 

\section{Oligomorphy and holography}
\label{sec:oligomorphs}

Having reviewed the basic features of cyclic and symmetric orbifolds, we now turn to the central question of interest: ``Can we delineate the subgroups $\ON < S_N$ which give rise to sparse low-lying spectra in the large $N$ limit?''. It turns to be useful to split the discussion into two parts. First in \S\ref{sec:cycle} we argue for a necessary and sufficient condition on the group $\ON$ to allow for a sensible large $N$ limit. Then  in \S\ref{sec:HartmanBound} we examine the constraints from modular invariance \cite{Hartman:2014oaa} and use it to give a sufficient condition for ensuring the sparseness of the low lying spectrum.

In order to motivate our conjecture, let us start by first examining the truncated partition function 
\eqref{eq:Z1truncated} where we have essentially kept the vacuum and the lowest primary state of weight 
$\Delta_1$. Let us ask how the untwisted spectrum of the orbifold theory looks like based on this data.\footnote{ One can run a similar argument for the lightest twisted sector primary. However, if the seed central charge $c$
is sufficiently large (we need $c > 8 \,\Delta_1$), it is clear that the leading contribution comes from the untwisted primary.} 
\begin{table}[t]
 \centering
{\tabulinesep=1.5mm
\begin{tabu}{|c||l||l|}
 \hline
      & seed theory  degeneracy: $d_1=1$ & orbit counting \\
  \hline
  $\mO(x^{\Delta_1})$ &$|\yes,\no,\no\rangle +|\no,\yes,\no\rangle + |\no,\no,\yes\rangle $ & $f_1^{(3)} = 1$ \\
  \hline
  $\mO(x^{2\Delta_1})$ &$|\yes,\yes,\no\rangle +|\no,\yes,\yes\rangle + |\yes,\no,\yes\rangle $ & $f_2^{(3)} = 1$ \\
  \hline
  $\mO(x^{3\Delta_1})$ &$|\yes,\yes,\yes\rangle $ & $f_3^{(3)} = 1$ \\
  \hline\hline 
       & seed theory degeneracy: $d_1 = 2$ & orbit counting \\ 
 \hline
  $\mO(x^{\Delta_1})$ & $|\ua,\no,\no\rangle +|\no,\ua,\no\rangle + |\no,\no,\ua\rangle $ & $d_1 f_1^{(3)} = 2\cdot 1 = 2$ \\
   & $|\da,\no,\no\rangle +|\no,\da,\no\rangle + |\no,\no,\da\rangle $  & \\
 \hline
  $\mO(x^{2\Delta_1})$ & $|\ua,\ua,\no\rangle +|\no,\ua,\ua\rangle + |\ua,\no,\ua\rangle $ & $d_1 f_2^{(3)}+\binom{d_1}{2}\, F^{(3)}_2 = 2\cdot 1 + 1\cdot 2 = 4$ \\
  & $|\da,\da,\no\rangle +|\no,\da,\da\rangle + |\da,\no,\da\rangle $  & \\
  & $|\ua,\da,\no\rangle +|\no,\ua,\da\rangle + |\da,\no,\ua\rangle$ & \\
  & $|\da,\ua,\no\rangle +|\no,\da,\ua\rangle + |\ua,\no,\da\rangle$ & \\
 \hline
\end{tabu}}
 \caption{Illustration of excited states in the $\Omega_3=\fZ_3$ orbifold of the seed theory with partition function $Z(x) = x^{-\frac{c}{12}}(1+d_1 x^{\Delta_1}+\ldots)$. We show multiple excitations of the same seed theory state with weight $\Delta_1$ in the untwisted sector; first for the case where the lowest lying state of the seed theory is non-degenerate and then for the case of it having degeneracy $2$. In the first case, the excitation is denoted by $\yes$ , in the second case the two modes of excitations are $\ua$ and $\da$. Every state of the orbifold theory is a linear combinations of states in which a definite choice of seed theory copies are excited, such that the linear combination is invariant under the $\fZ_3$ action. The right column shows how the number of orbifold states can be obtained from counting orbits $f_\ell^{(3)}$ and $F_\ell^{(3)}$.}
 \label{table:oligomorphic}
 \end{table}
First, given a unique seed vacuum, the orbifold $\Cperm$ has again a unique vacuum, clearly obtained by acting with the identity on all the $N$-copies and projecting out by $\ON$. For the first excited state, we can act with ${\cal O}_{\Delta_1}$ on any of the $N$-copies to construct a state. But requiring that the resulting state be $\ON$ invariant, we are meant to consider the orbits of $\ON$ on the set $X_N$.  If $\ON$ is transitive, i.e. for every ${j,k} \in X_N$ there exists a $z \in \ON$ such that $z\cdot j = k$, then there is a single orbit under the $\ON$ action, as we can cover all the $N$  copies of ${\cal C}$ by acting with a symmetry generator. If $\ON$ acts intransitively, then we may have more invariant states measured by the number of orbits of $\ON$. Letting the number of orbits of $\ON$ be $f_1^{(N)}$ we have the degeneracy of the first excited state to be $d_1\, f_1^{(N)}$. We will get similar contributions from other primaries.

Next let us estimate the degeneracy of the state with weight $2\Delta_1$ which is obtained by acting twice with 
${\cal O}_{\Delta_1}$. To excite the seed theory state with weight $\Delta_1$ twice, ${\cal O}_{\Delta_1}$ has to act on two different copies. The resulting state then has to be superposed with all other states in the same orbit of $\ON$ such that an $\ON$-invariant orbifold state is obtained. Let us now account for non-trivial degeneracy $d_1$: every action of ${\cal O}_{\Delta_1}$ excites one out of $d_1$ possible modes. Assume first both excitations are excitations of the same mode. Then, after summing over images of these excitations under $\ON$, the number of different orbifold states of this type is given by the number $f_2^{(N)}$ of orbits of $\ON$ on unordered pairs of $2$ elements of $X_N$. This number has to be multiplied by $d_1$ to account for the different possible modes of the excitations: $d_1 f_2^{(N)}$. Next, consider orbifold states where ${\cal O}_{\Delta_1}$ excites two different modes. Since the two excitations are now distinguishable, the number of such states is  given by $F_2^{(N)}$, i.e., the number of orbits of $\ON$ on ordered $2$-tuples of distinct elements of $X_N$.\footnote{ The elements need to be distinct because ${\cal O}_{\Delta_1}$ cannot excite the same state twice on the same copy.} Again, the result has to be multiplied by a factor $\binom{d_1}{2}$ to account for the different combinations of modes that could be excited this way: $\binom{d_1}{2} F^{(N)}_2$. See table \ref{table:oligomorphic} for an example of this counting.

This can nicely be formalized in terms of properties of the group $\ON$. For a degree $N$ permutation group acting on a  set $X_N$, we see that there is a natural action on $\ell$-element subsets which is induced in addition to an action on $\ell$-tuples of distinct elements (corresponding to the two situations described above).
Let
\begin{itemize}
\item $f_\ell^{(N)}$ be number of orbits of $\ON$ on (unordered) $\ell$-element subsets of $X_N$,
\item $F_\ell^{(N)}$ be the number of orbits of $\ON$ on (ordered) $\ell$-tuples of distinct elements from $X_N$.
\end{itemize}
In general, the number of $\mO(x^{\ell \Delta_1})$ states obtained from acting $\ell$ times with $\mO_{\Delta_1}$ is a polynomial function of $f_\ell^{(N)}$, $F_\ell^{(N)}$ which depends also on $d_1$. 

In the following we will sometimes be concerned with properties (such as existence) of the limits $\lim_{N\rightarrow\infty} f_\ell^{(N)}$ and $\lim_{N\rightarrow\infty} F_\ell^{(N)}$. In such contexts we will always work under the following assumptions.\footnote{  We thank an anonymous referee for pointing out to us the necessity of these assumptions.} The limits above will be assumed to refer to a family of permutation groups $\{\ON\}_{N\in \,\mathcal{I}}$ where $\mathcal{I} \subset \fZ_+$ is some infinite set of integers to index the groups. Further the groups $\ON$ of such a family shall be such that if the numbers $f_\ell^{(N)}$ and $F_\ell^{(N)}$ are bounded as $N\rightarrow \infty$, then they do converge to definite limiting values denoted as $f_\ell \equiv \lim_{N\rightarrow\infty} f_\ell^{(N)}$ and $F_\ell\equiv\lim_{N\rightarrow\infty} F_\ell^{(N)}$, respectively. This requirement is just formalizing the fact that we want the members of a family of permutation groups to define CFTs in an operationally equivalent way, but at increasing values of central charge (e.g., the groups $\{S_N\}_{N\in\fZ_+}$ are comparable because their action is always defined in the same way for any $N$, just on a different number of copies of the CFT).

We are now in a position to state the criterion for $\Cperm$ to have a universal low lying spectrum and to admit a stringy holographic dual. Clearly a primary requirement is the finiteness of the degeneracies of low lying states in the orbifold CFT partition function as $N\rightarrow \infty$. We will show that this is equivalent to the requirement that $f_\ell^{(N)}$ for $\ell\ll N$ remain finite. In case a limiting permutation group $\Oinf \equiv \lim_{N\rightarrow \infty} \ON$ exists, this criterion is precisely what defines {\em oligomorphic permutation groups} \cite{Cameron:1990fk}. The following subsection is devoted to making this statement more precise. However, finiteness by itself doesn't guarantee that we will have a spectrum that conforms to holographic expectations. Therefore \S\ref{sec:HartmanBound} we will turn to stating an additional criterion about the precise growth rate of $f_\ell$ with $\ell$. According to \cite{Hartman:2014oaa} the growth should be at most exponential in order to get a theory with holographic dual. This ends up carving out a subspace of permutation groups with bounded $f_\ell$, which we argue is sufficient.

\subsection{P\'olya counting for the partition sum}
\label{sec:cycle}

To formalize the statements, we now need some information about the structure of permutation groups. 
As we saw above, the crucial piece of data we require is the number of orbits of $\ON$ on $\ell$-element subsets 
 of $X_N$. This can as always be formalized into a generating function, but first we need some essential information from P\'olya counting theory about the structure of the permutation group $\ON$.\footnote{ We would like to thank Alex Maloney for a discussion on this issue.} 

Recall that every element $y \in \ON$ being a permutation can be viewed as an element of $S_N$ and thus admits a cycle decomposition. As before we will refer to $\{m_k\}_N$ as the cycle type of $y$. We define the cycle index of $\ON$ as the group averaged representation of the cycle types of its elements. More specifically, given a set of variables 
$\gamma_i$, with $i \in X_N$, 
\begin{align}
y  = (1)^{m_1}\, (2)^{m_2}\, \cdots (N)^{m_N} \;\; \mapsto \;\;
\gamma_1^{m_1}\,\gamma_2^{m_2}\cdots\gamma_N^{m_N} \,.
\label{eq:inmon}
\end{align}
The cycle index for $y \in \ON$ is a monomial in the $\gamma_i$ encoding its cycle type and averaging this over all elements we get the cycle index $\cindex$ of $\ON$. To wit,
\begin{equation}
\cindex(\ON; \gamma_1, \gamma_2, \cdots , \gamma_N) = \frac{1}{|\ON|}\, \sum_{y \in \ON}\,
\prod_{i=1}^N\, \gamma_i^{m_i(y)}
\label{}
\end{equation}	
where we have retained the explicit $y$-dependence in the $m_i(y)$ for clarity of notation.

Given the cycle index for a group it is easy to work out the numbers $f_k^{(N)}$ and $F_k^{(N)}$. These are given by what are sometimes called the ordinary generating function and the exponential generating function respectively and are defined by choosing specific values of $\gamma_i$ in terms of the generating parameter:
\begin{align}
\sum_{\ell=0}^N \, f_\ell^{(N)}\, t^\ell  &= \cindex\left(\ON; 1+ t, 1+t^2 , \cdots,  1+t^N \right) \,, 
\qquad \gamma_\ell = 1+t^\ell
\nonumber \\
\sum_{\ell=0}^N \, \frac{1}{\ell!} \, F_\ell^{(N)}\, t^\ell  &= \cindex\left(\ON; 1+ t, 1,\cdots ,1 \right) 
\,,\quad \gamma_\ell = 1+ t^\ell \delta_{\ell1}
\label{eq:fFdef}
\end{align}

Armed with this group theoretic information let us revisit the orbifold partition function 
\eqref{eq:Z1general}. We recognize a similar structure there, which is of course no coincidence. In constructing the orbifold partition sum we sum over commuting elements  \
$z_a, z_b \in \ON$. We can decompose this sum  by first restricting attention to $z_a = e$ whence all values of $z_b$ are allowed, and then estimating the contribution from non-trivial elements $z_a$. In the first instance the joint orbits $\mO(z_a,z_b)$ are determined by the cycle type of $z_b$. In fact, from \eqref{eq:Z1general} we see immediately the connection between cycle index and the untwisted sector of orbifold CFT partition functions: 
\begin{align}
Z_{\perm}(\tau,\btau) &= \frac{1}{|\ON|} \sum_{\substack{z_a,z_b\in \ON \\ z_a z_b = z_b z_a}}  \ 
 \prod_{\xi \in \mO(z_a,z_b)} Z(\tau_\xi,\btau_\xi) \nonumber \\
&= \cindex\left(\ON; Z(\tau, \btau), Z(2\, \tau, 2\,\btau), \cdots, Z(N\tau, N\btau) \right)
\nonumber \\
& \qquad +\;
\frac{1}{|\ON|}\;\sum_{\substack{z_a , z_b\in \ON \\ z_a \neq e \;\, \\ z_a z_b = z_b z_a }}  \ 
 \prod_{\xi \in \mO(z_a,z_b)} Z(\tau_\xi,\btau_\xi) 
  \,.
\label{eq:CycleIndexIdent}
\end{align}
The first line of the final expression includes contributions from $z_a = e, z_b\in \ON$ for which the product over orbits $\mO(z_a,z_b)=\mO(z_b)$ becomes a product of the form as it appears in the definition of the cycle index. As we will show shortly, this captures the entire untwisted sector and thus the large $N$ spectrum of the orbifold theory in the low temperature regime. In the high temperature regime a similar expression would dominate where the insertions $Z(k\tau,k\btau)$ in the cycle index would be replaced by $Z\left(\frac{\tau}{k},\frac{\btau}{k}\right)$, corresponding to a modular transformation of the expression in \eqref{eq:CycleIndexIdent}.\footnote{ Note in particular that if instead we isolate the $z_b = e$ contribution to $Z_{\perm}(\tau,\btau)$ we would get the cycle index $\cindex\left(\ON; Z(\tau, \btau), Z(\frac{ \tau}{2},\frac{\btau}{2}), \cdots, 
Z(\frac{\tau}{N}, \frac{\btau}{N}) \right) $.}  

In the following we want to argue that if we can control the cycle index expression in 
\eqref{eq:CycleIndexIdent}, then the essential properties of the full $Z_{\perm}(\tau,\btau)$ follow. In particular, once we are able to argue that the first line has holographic properties, then so does the full $Z_{\perm}(\tau,\btau)$. Let us first argue this for the example of cyclic and symmetric orbifold theories by showing that the first line of (\ref{eq:CycleIndexIdent}) is enough to reproduce the results of Theorems \ref{thm:1} and \ref{thm:2}. Afterwards we will turn to a derivation for generic $\ON$. 

\paragraph{Cycle index for $\fZ_N$ orbifolds:}
In the case of cyclic orbifolds it is easy to see that the cycle index evaluated on the seed partition function captures the large-$N$ dominant part. To this end observe that for any $y\in \ON = \fZ_N$ it holds $m_i(y) = (N,y)$ for $i = \frac{N}{(N,\,y)}$ and $m_i(y) = 0$ for all other indices $i$. From this it follows that
\begin{align}
 &\cindex\left(\fZ_N; Z(\tau, \btau), Z(2\, \tau, 2\,\btau), \cdots, Z(N\tau, N\btau) \right) \notag\\ 
   &\qquad= \frac{1}{N} \sum_{y\in \fZ_N} \;\prod_{k=1}^N Z(k\tau,\, k\btau)^{m_k(y)} \notag\\
   &\qquad= \frac{1}{N}\left[ Z(\tau,\btau)^N + \sum_{y=1}^{N-1} Z\left( \frac{N}{(N,y)} \tau,\, \frac{N}{(N,y)} \btau \right)^{(N,y)} \right] \,,
\end{align}
which is precisely the part of the full $\fZ_N$-orbifold partition function which is relevant at large $N$ according to Theorem \ref{thm:1}. 

\paragraph{Cycle index for $S_N$ orbifolds:}
As we have seen in  \S\ref{sec:freeb} and \S\ref{sec:asycs}, the large-$N$ partition function $Z_{\sym}(x)$ has a finite low lying spectrum which depends on the seed theory. By restricting just to the untwisted sector described by the cycle index in the first line of (\ref{eq:CycleIndexIdent}) we will certainly be unable to capture all of these low lying states. However, because their degeneracy is finite as $N\rightarrow \infty$, they become irrelevant at large $N$. In order to determine the large $N$ asymptotics of $Z_{\sym}(x)$ it is thus sufficient to reproduce the highly degenerate spectrum of heavy operators. This is indeed captured by the cycle index as one can argue by looking at
\begin{align}\label{eq:CycleIndexS}
 &\cindex\left(S_N; Z(\tau, \btau), Z(2\, \tau, 2\,\btau), \cdots, Z(N\tau, N\btau) \right) \notag\\ 
 &\qquad= \frac{1}{N!} \sum_{y\in S_N} \; \prod_{k=1}^N Z(k\tau,\, k\btau)^{m_k(y)} \notag\\
 &\qquad= \sum_{\{m_k\}_N} \; \prod_{k=1}^N \frac{1}{m_k!} \, \left[T_k^{\text{(trc)}}Z(\tau,\, \btau)\right]^{m_k} \,,
\end{align}
which is just the expression (\ref{eq:GeneralSnresult}) for the $S_N$ orbifold partition function with truncated Hecke operators
\begin{align}\label{eq:HeckeTruncate}
 T_{k}^{\text{(trc)}} Z(\tau,\btau) \equiv 
 \frac{1}{k} \left[\sum_{d|k} \;\sum_{\kappa=0}^{d-1} Z\left(\frac{k \tau + \kappa d}{d^2},\,\frac{k \btau + \kappa d}{d^2}\right) \right]_{d=1}
 = \frac{1}{k} \, Z(k\tau,\, k\btau) \,.
\end{align}
This means that the cycle index (\ref{eq:CycleIndexS}) captures precisely the part of the $S_N$ orbifold partition function which arises from replacing every Hecke operator by its large-$N$ dominant contribution as in (\ref{eq:HeckeTruncate}). According to Theorem \ref{thm:0} (or more quantitatively as in Appendix \ref{sec:excitedstates}), at large $N$ the relevant contributions from a Hecke operator are given by its truncated piece up to irrelevant corrections:
\begin{align}
T_N Z(x) = T_N^{\text{(trc)}} Z(x) + x^{-\frac{\cc}{12}}\; \mO \left( x^{\geq \frac{\cc}{16}} \right) \,.
\label{eq:htrn}
\end{align}
Therefore, (\ref{eq:CycleIndexS}) is the only relevant piece at large $N$ and captures all the properties of $Z_{\sym}(\tau,\btau)$ which give it the holographic universality. 

\paragraph{Cycle index for general orbifolds:}
Let us now consider the case of arbitrary permutation group $\ON$. By the same reasoning as before, the cycle index in the first line of (\ref{eq:CycleIndexIdent}) captures the large-$N$ behavior of $Z_{\perm}(\tau,\btau)$. We should therefore be able to find a criterion for whether or not a given $\ON$ leads to an orbifold theory with stringy holographic dual by simply examining the cycle index. Indeed this has all the necessary ingredients to tell us whether or not the counting of low-lying states is commensurate with our expectations.
We therefore claim the following:
\begin{theorem}
The orbifold theory $\Cperm$ has degeneracies of low lying states which remain finite as $N\rightarrow \infty$ if and only if $f_\ell^{(N)}$ stays finite for all $\ell$ in this limit. In situations where a limiting group $\Oinf$ exists, this statement is equivalent to $\Oinf$ being oligomorphic.
\end{theorem}

A simple argument shows that finiteness of $f_\ell$ is a sufficient condition for having a universal low lying spectrum. Consider the large $N$ leading contribution of the seed theory $Z(x) = x^{-\frac{c}{12}} \sum_k \, d_k \, x^{\Delta_k}$ to the full permutation orbifold partition function: 
\begin{align}
Z_{\perm}(\tau,\btau) &\simeq 
\cindex\left(\ON; Z(\tau, \btau), Z(2\, \tau, 2\,\btau), \cdots ,Z(N\tau, N\btau) \right) \nonumber \\
&= x^{-\frac{\cc}{12}} \left[ \frac{1}{|\ON|} \sum_{y\in\ON}\, \prod_{k=1}^N\; \left(1+d_1\, x^{k\Delta_1}+d_2\, x^{k\Delta_2}+\ldots\right)^{m_k(y)}\right]  \notag\\
&\geq x^{-\frac{\cc}{12}} \left[ \frac{1}{|\ON|} \sum_{y\in\ON}\, \prod_{k=1}^N\; \left(1+ x^{k\Delta_1}\right)^{m_k(y)}\right] \notag\\
&= x^{-\frac{\cc}{12}} \, \sum_{\ell=0}^N f_\ell^{(N)}\; x^{\ell\Delta_1}  \,,
\label{eq:inequ1}
\end{align}
which means that if $Z_{\perm}(\tau,\btau)$ has degeneracies of low lying states which are $N$-independent for $\ell\ll N$, then $f_\ell^{(N)}$ must not grow with $N$ either (for $\ell\ll N$).

On the other hand, in order to show the necessity of the criterion about finiteness of $f_\ell$, it is, of course, enough if we restrict to contributions from the low-lying states of the seed theory, i.e., 
\begin{align}
Z(x) = x^{-\frac{c}{12}} \left( 1 + \sum_{k=1}^K d_k \, x^{\Delta_k} + \ldots \right) \,,
\label{eq:SeedTrunc}
\end{align}
where $K\equiv K(N)$ is chosen such that $\tilde{d}_K \equiv \sum_{k=1}^K d_k \ll N$ and $\Delta_K \ll N \Delta_1$; in this sense ellipses denote higher order terms that we can discard for an analysis of the low lying spectrum.  
The leading contribution to $Z_{\perm}(x)$ at large $N$ is then
\begin{align}
Z_{\perm}(x) 
&\simeq x^{-\frac{\cc}{12}} \left[ \frac{1}{|\ON|} \sum_{y\in\ON}\, \prod_{j=1}^N\; \left(1+ \sum_{k=1}^K d_k \, x^{j\Delta_k}+\ldots\right)^{m_j(y)}\right]  \notag\\
&\leq x^{-\frac{\cc}{12}} \left[ \frac{1}{|\ON|} \sum_{y\in\ON}\, \prod_{j=1}^N\; \left(1+ \tilde{d}_K \, x^{j \Delta_1}+\ldots\right)^{m_j(y)}\right]  \notag\\
&\leq x^{-\frac{\cc}{12}} \left[ \frac{1}{|\ON|} \sum_{y\in\ON}\, \prod_{j=1}^N\; \left(1+ (\tilde{d}_K \, x^{\Delta_1})^j +\ldots\right)^{m_j(y)}\right]  \notag\\
&= x^{-\frac{\cc}{12}} \,\sum_{\ell=0}^N (\tilde{d}_K)^\ell\, f_\ell^{(N)} \; x^{\ell\Delta_1} + \ldots \,,
\label{eq:ineq2}
\end{align}
where we used (\ref{eq:fFdef}). For $\ell\ll N$ the interesting behavior of the degeneracy factor $(\tilde{d}_K)^\ell\, f_\ell^{(N)}$ is determined by $f_\ell^{(N)}$ since $(\tilde{d}_K)^\ell$ are finite by definition. From this we can now explicitly see that a necessary condition for a freeze-out of low lying states is $N$-independence of 
$f_\ell^{(N)}$ for $\ell\ll N$, i.e., finiteness of $f_\ell$. 

Groups $\Oinf$ of unbounded degree acting as permutations on the infinite set $X_\infty \simeq {\mathbb Z}_+$, such that for every natural number $\ell$, $\Oinf$ has only finitely many orbits on $(X_\infty)^\ell$  are said to be {\em oligomorphic} \cite{Cameron:1990fk,Cameron:2009rr}. Therefore finiteness of the low-lying states is tantamount to the requirement that the  group $\ON$ limits to an oligomorphic permutation group.

\subsection{Bound on growth rate of holographic CFT spectra}
\label{sec:HartmanBound}
We have seen that a necessary and sufficient condition for an $N$-independent low lying spectrum of $\Cperm$ is the $N$-independence of $f_\ell^{(N)}$ for $\ell\ll N$. However, this by itself does not guarantee that the resulting orbifold CFT has a string holographic dual in the sense described in \S\ref{sec:introduction}. The precise criterion we need has recently been derived in \cite{Hartman:2014oaa}, exploiting the constraints from modular invariance. Using the upper bound presented there for density of states of the low lying primaries we can further constrain to a subset of oligomorphic permutation groups $\Oinf$, by bounding the growth rates of $f_\ell$.

To wit, consider a CFT with parametrically large central charge $\cc=Nc$ and partition function  
\begin{align}
  Z_{\perm}(x) = \sum_{E\geq -\frac{\cc}{12}} \rho(E) \, x^E = x^{-\frac{\cc}{12}} \sum_{E\geq 0} \rho\left(E-\frac{\cc}{12}\right) \, x^E \,.
\end{align}
According to \cite{Hartman:2014oaa} this CFT has a (stringy) holographic dual provided that
\begin{align}
\rho\left(E-\frac{\cc}{12}\right) \lesssim e^{2\pi E} \,.
\end{align}
From the upper bound on orbifold state degeneracies that we derived in (\ref{eq:ineq2}), we can immediately infer that for our class of theories, we have
\begin{align}
 x^{-\frac{\cc}{12}}\sum_{\ell=0}^N f_\ell^{(N)} \, x^{\ell\Delta_1} \leq x^{-\frac{\cc}{12}} \sum_{E\geq 0} \rho\left(E-\frac{\cc}{12}\right) \, x^E \,,
\end{align}
From this we can give a necessary criterion on $\Omega_N$ for it to give rise to a holographic orbifold theory: 
\begin{theorem} \label{thm:oligo}
 In order for the permutation orbifold $\Cperm$ to have a holographic dual in classical string theory it is a necessary that not only $f_\ell^{(N)}$ be independent of $N$ for $\ell \ll N$ (i.e., $\Omega_\infty$ if well-defined is oligomorphic) but in addition the growth in $\ell$ is at most exponential:
 \begin{align}
   f_\ell^{(N)} \lesssim e^{2\pi \, \ell \,\Delta_1} \,, \qquad {\rm for}\;\; \ell \ll N \,,
   \label{eq:GrowthBound}
 \end{align}
where $\Delta_1$ is the energy gap in the seed theory ${\cal C}$. 
\end{theorem}

This completes our analysis of the criteria required for the permutation orbifold to behave `matrix-like' in the large $N$ limit. While the situation is well exemplified by the symmetric orbifolds as we have reviewed earlier, for the reminder we focus on outlining other examples that satisfy the criteria of Theorem \ref{thm:oligo}.

\section{Oligomorphic permutation groups: examples and growth rates}
\label{sec:oligomorphic}

Oligomorphic permutation groups have been studied extensively in the literature. We will just mention some of the salient features, to be found in \cite{Cameron:1990fk}, which are relevant to our discussion. 

First of all it is important to distinguish between finite permutation groups $\ON$ acting on $X_N$ and their limit $\Oinf$ acting on $X_\infty = \fZ_+$. Finite groups are trivially oligomorphic, so the interesting oligomorphic groups are permutation groups of the second kind for which the number of orbits on $\ell$-element sets of positive integers stays finite for all $\ell$. Note that the number of infinite order permutation groups is vast, e.g., $S_\infty$ (the group of permutations on positive integers) has an uncountably infinite order. It is therefore important to know that even for uncountable permutation groups acting on $\fZ_+$ there is always a subgroup of countable order which has the same $F_\ell$ characteristics.\footnote{ A proof of this rather remarkable statement can be found in \cite{Cameron:1990fk} where it is stated to be a consequence of a mapping between permutation groups and model theory and relies on a theorem by 
L\"owenheim and Skolem in the latter.} For our analysis we can thus always restrict ourselves to permutation groups of countable order.

Oligomorphic permutation groups are typically exemplified by the following:
\begin{itemize}
\item  $S_\infty$: the symmetric group of infinite order acting on $X_\infty$
\item $A_\infty$: the group of order preserving permutations of rationals ${\mathbb Q}$
\end{itemize}
The former of course naturally arises as the large $N$ limit of the regular symmetric groups $S_N$ acting on finite sets, but the latter appears to make sense only as an oligomorphic  group with no finite $N$ analog. 

These two groups are highly homogeneous but only $S_\infty$ is highly transitive, where one defines these concepts based on the $\ell$ dependence of $f_\ell$ and $F_\ell$ respectively, viz.,
\begin{itemize}
\item  highly homogeneous  \; $\Longleftrightarrow \;\; f_\ell =f_{\ell+1} \; \forall \; \ell$ 
\item  highly set-transitive  \; $\Longleftrightarrow\;\ f_\ell = 1 \; \forall \; \ell$ 
\item  highly transitive \; $\Longleftrightarrow \; \; F_\ell =1 \; \forall \; \ell$ 
\end{itemize}
For the group $A_\infty$, $F_\ell  = \ell!$. The highly homogeneous property of $A_\infty$ leads us to suspect that there might be a stringy holographic dual in the strict limit. However, we would like to argue that the absence of a family of finite $N$ (thus finite central charge) theories which limit to the formally ${\cal C}^{\otimes\infty}/A_\infty$ theory, makes this an uninteresting example for physical purposes. We will shortly argue for a more interesting class of theories based on a simple group theoretic construction.\footnote{ It is amusing to note that if we focus on permutation groups of finite order and degree $N$ and require that the action of the group be highly set-transitive (i.e., it act transitively on all $X_k \subset X_N$ for $k\leq N$), then the only such groups for $N \geq 6$ are the symmetric and alternating groups \cite{Beaumont:1955kx,Livingstone:1965fk}, which can be argued from the classification theorem of finite simple groups. The Mathieu group $M_{24}$ is the only other group which is 5-set transitive.}

While having a highly homogeneous oligomorphic permutation group would do the job, the number of these with countable degree is quite restrictive; \cite{Cameron:1990fk} argues for such groups being a simply a dense subgroup of $S_\infty$, $A_\infty$,  the group $C_\infty$ (permutations of roots of unity preserving cyclic order) and two others groups $B_\infty$ and $D_\infty$ (which allow order reversal in $A_\infty$ and $C_\infty$ respectively), cf., \cite{Cameron:1990fk,Cameron:1999ys}. By the logic above none of these are interesting from a holographic perspective. Fortunately for us, we need not impose a condition as strong as highly homogeneous; Theorem \ref{thm:oligo} only requires that the growth of $f_\ell$ be not too fast. This does allow the presence of other oligomorphic groups some of which occur in families admitting finite $N$ analogs.

Before we discuss explicit examples however, let us record one interesting fact about the growth rate of the $f_\ell$ for oligomorphic permutation groups. To do so, we need one extra notion of {\em primitivity}. A permutation group $\Oinf$ is said to be primitive if its action on $X_\infty$ has only the trivial equivalence relation (elements being equivalent to themselves) and the universal equivalence relation (the set being equivalent to itself). Given this notion, Theorem 4.1 of \cite{Cameron:2009rr} asserts that for primitive, but not highly set-transitive  $\Oinf$, $f_\ell \geq n^\ell/p(\ell)$ for some constant $n$ and polynomial $p(x)$. In particular, it is argued that there is a gap in the growth of $f_\ell$ between a constant and exponential. The former would be desirable for us given \eqref{eq:GrowthBound}.  When the $f_\ell$(s) grow exponentially we would have to examine the situation more closely, since our tolerance on the rate of growth depends on the details of the seed theory ${\cal C}$ through $\Delta_1$. 
We note in passing that there is no upper bound on the growth rate of $f_\ell$ among the oligomorphic groups, so clearly not all oligomorphic groups will satisfy our criteria and give rise to stringy holographic duals.

As mentioned above,  for the physical application to large central charge CFTs, we are interested in oligomorphic groups which smoothly connect to finite degree permutation groups $\ON$. Demanding at most exponential growth of $f_\ell^{(N)}$ as a function of $\ell$ for $\ell\ll N$, we believe still leaves a large class of examples to explore. For instance, from the list of examples and explicit growth rates for various oligomorphic groups that can be found  in \cite{Cameron:2000zr}, one can see that oligomorphic groups with growth rates of $f_\ell$ faster than exponential are typically somewhat exotic. It is tempting to speculate that using the physical criterion of Theorem \ref{thm:oligo} one can demarcate the class of oligomorphic permutation groups further (e.g, the gap in the growth rate  mentioned above is suggestive).

While the examples we have described above are quite exotic, it is possible using standard group theoretic constructions to  conjure a wide class of interesting physical examples. For instance we can, of course, take direct product groups, but it is more interesting to exploit the wreath product construction which leads to a plethora of permutation orbifolds as we describe below in \S\ref{sec:wreath}. 

\section{Wreath product orbifolds}
\label{sec:wreath}

We now turn to interesting examples of permutation orbifolds which we obtain using the wreath product construction in group theory.  The theories thus obtained which we will call wreath product orbifold CFTs are interesting for several reasons. First of all, they provide a way to construct a rich class of holographic permutation orbifolds where the order of $\ON$ is between that of $\fZ_N$ and $S_N$. In addition, they are actually rather natural constructs in computation of entanglement entropy for CFTs.  To see this, consider the computation of  R\'enyi entropies for permutation orbifold CFTs in two dimensions. The replica method which is used as a technical tool to achieve this, involves considering a further cyclic orbifold by, say, ${\mathbb Z}_q$ (to compute the $q^{\rm th}$ R\'enyi entropy). This secondary cyclic orbifolding can be combined with the permutation action we had in theory via the wreath product. More precisely,  the $q^{\rm th}$ R\'enyi entropy of the $p$-fold symmetric product orbifold ${\cal C}^{\otimes p} / S_p$ is determined by certain correlation functions in the wreath product orbifold CFT ${\cal C}^{\otimes pq} / (S_p \wr \fZ_q)$. The group $S_p \wr \fZ_q$ has degree $pq$ and order $q(p!)^q$ and is obtained by the wreath product construction.\footnote{ We adopt the standard notation $\wr$ for the wreath product.}

\subsection{Wreath products of permutation groups}
\label{sec:WreathPerm}

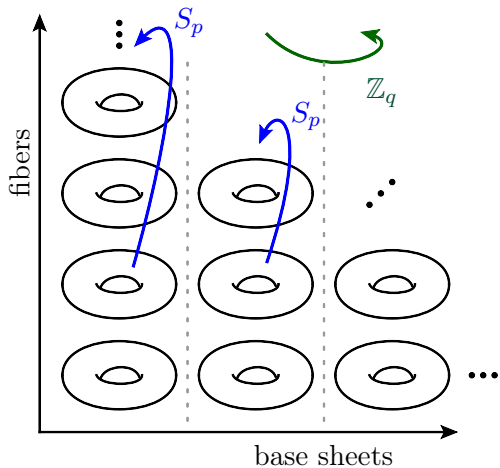
\begin{figure}
 \centerline{\includegraphics[width=.4\textwidth]{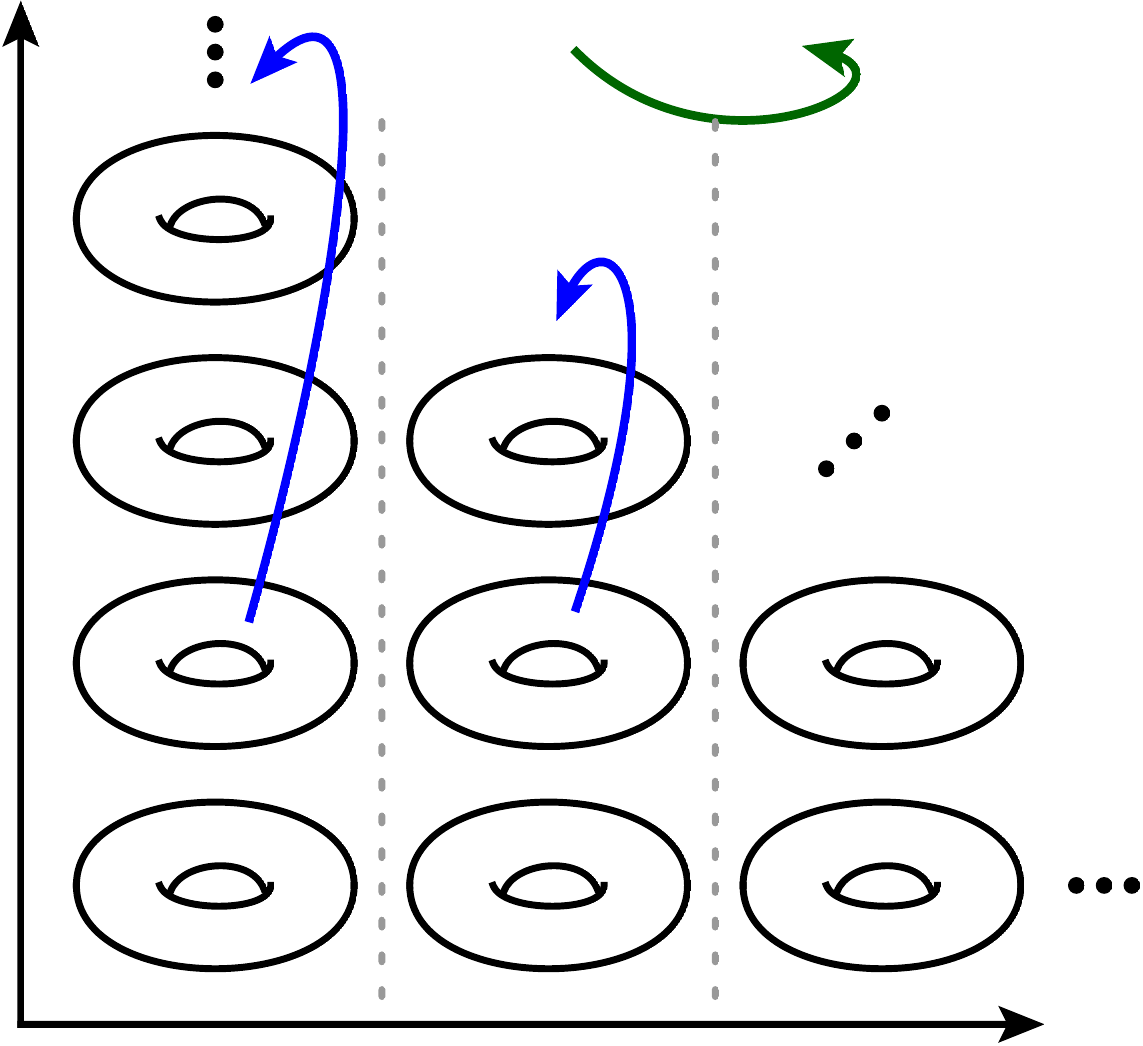}}
  \setlength{\unitlength}{0.1\columnwidth}
 \begin{picture}(0.3,0.4)(0,0)
 \put(5.35,3.2){\makebox(0,0){\textcolor{blue}{$S_p$}}}
 \put(4.35,4){\makebox(0,0){\textcolor{blue}{$S_p$}}}
 \put(6,3.4){\makebox(0,0){\textcolor{green}{$\fZ_q$}}}
 \put(5.5,.3){\makebox(0,0){base sheets}}
 \put(2.9,2.8){\begin{turn}{90}\makebox(0,0){fibers}\end{turn}}
 \end{picture}
 \caption{Illustration of the $S_p \wr \fZ_q$ wreath product action on $pq$-sheeted covers of a torus. The cyclic symmetry $\fZ_q$ acts on a set of $q$ base sheets of the parent torus. This set of $q$ tori is fibred where each fiber consists of a $p$-sheeted unbranched cover. The symmetric group $S_p$ acts independently on each fiber. The imprimitive wreath product action preserves the fibration, i.e., it preserves the presence in each column of one copy of each orbifold sheet.
 }
 \label{fig:wreath}
 \end{figure}

Let us start by reviewing the wreath product construction. Consider two permutation groups, $G_p$ and $H_q$, acting on $X_p$ and $X_q$ respectively. For definiteness, we will define $N=pq$ such that wreath products of $G_p$ and $H_q$ are subgroups of $S_N$. Using the wreath product we can construct all groups that act on $X = (X_p)^{q}$ and preserve the partitioning of $X$ into factors of $X_p$ (clearly the cardinality of $X$ is $N$). We are particularly interested in the unrestricted wreath product of $G_p$ and $H_q$ themselves:
\begin{align}
  G_p \wr H_q = G_p^q \rtimes H_q &\equiv \{ (g_1,\ldots,g_q,\sigma) \; : \; g_i \in G_p, \; \sigma \in H_q \} \,, \\
  \text{with}\quad (g_1,\ldots,g_q,\sigma) \circ (\bar g_1, \ldots, \bar g_q, \bar \sigma) &=
       (g_1 \bar g_{\sigma^{-1}(1)} , \ldots , g_q \bar g_{\sigma^{-1}(q)}, \sigma \bar \sigma ) \,, \notag\\
       (g_1,\ldots,g_q,\sigma)^{-1} &= (g^{-1}_{\sigma(1)} , \ldots, g^{-1}_{\sigma(q)}, \sigma^{-1} ) \,,
\end{align} 
where the last two lines define the group action and the inverse. There is a simple pictorial interpretation of the wreath product: the set $X$ that $G_p\wr H_q$ acts on can be visualized as an $p\times q$ matrix $M$ where each column contains the numbers $1,\ldots,p$. An element  $(g_1,\ldots,g_q,\sigma) \in G_p\wr H_q$ acts on this matrix by first permuting the numbers within each column as prescribed by the $g_i$, i.e., $M_{i,j} \mapsto M_{g_j(i),\,j}$. Then the columns are permuted by $\sigma$, i.e., $M_{g_j(i),\,j} \mapsto M_{g_{\sigma(j)}(i),\,\sigma(j)}$. From this picture we can easily see that the action of $G_p \wr H_q$ on $\Omega$ is transitive and imprimitive, i.e., it preserves the decomposition of the matrix into columns that contain the numbers $1,\ldots,p$. Using language adapted to partition functions (and covers of tori), a simple way of saying this is that we have a set of $q$ replica copies which are fibred. Each fiber contains another $p$ copies of the torus and $S_p$ acts independently on each fiber. See Fig.~\ref{fig:wreath} for an illustration.

Since the order of the group is the first intuitive indicator for whether or not the corresponding orbifold CFTs have a universal holographic spectrum, let us briefly note the order of wreath product groups, which can be quite large, for 
\begin{align}
  |G_p \wr H_q| = |G_p|^q \, |H_q| \,. 
\end{align}

Armed with this information let us  flesh out the connection between entanglement and wreath product constructions, which we have alluded to earlier. If we want to understand the $q^{\rm th}$ R\'enyi entropy in the $S_p$ symmetric product orbifold theory, we are instructed to compute certain correlation functions in a replica geometry which has an additional $\fZ_q$ symmetry. The geometric picture that we described above resembles exactly the structure of a $q$-fold replica cover of the $p$-fold cover that is used to describe the permutation orbifold. The wreath product action implements a cyclic symmetry on the replica surface base space. At the same time, it gives a fibration of this base, with each fiber being a tower of $p$ orbifold sheets on which an independent $S_p$ acts. The CFT whose partition function encodes the $q^{\rm th}$ R\'enyi entropy of the $S_p$ symmetric product orbifold theory ${\cal C}^{\otimes p}/S_p$ is therefore just ${\cal C}^{\otimes pq}/(S_p \wr \fZ_q)$ with central charge $\cc\equiv Nc=pqc$.\footnote{ We are simply giving an illustrative picture here to draw attention to the connection to R\'enyi entropies. To actually compute entanglement entropy, we need to specify a region etc., which will involve introducing branching points (where twist operators) are inserted etc..}

\subsection{Partition functions}
\label{sec:wreathpartitions}

The primary motivation for us to study wreath products is the rich landscape of tractable orbifold models that they give rise to. This is due to the fact that it is very easy to generate partition functions of wreath product orbifold CFTs: start with some CFT ${\cal C}$ with central charge $c$. From the remarks in the previous subsection it is not hard to verify that the $(G_p\wr H_q)$-orbifold of ${\cal C}$ has central charge $\cc=Npq$ and it is obtained by performing two subsequent orbifold projections, first with respect to $G_p$, then with respect to $H_q$, i.e.,
\begin{align}
{\cal C}^{\otimes pq} / (G_p\wr H_q) = \left({\cal C}^{\otimes p}/ G_p \right)^{\otimes q} / H_q \,.
\label{eq:WreathIdent}
\end{align}
This identity has a profound manifestation at the level of the cycle index of wreath product groups \cite{Cameron:1990fk}: 
\begin{align}
\cindex(G_p\wr H_q;\, \gamma_1, \gamma_2, \cdots , \gamma_N) = \cindex(H_q;\, \eta_1,\eta_2,\cdots,\eta_q)\,,
\end{align}
where $\eta_k = \cindex(G_p;\, \gamma_k,\gamma_{2k},\cdots,\gamma_{pk})$.
Let us now mention two of the most straightforward examples of wreath product orbifolds.

\paragraph{$S_p \wr \fZ_q$ orbifolds:}
\label{sec:WreathSpectra1}
Consider an orbifold CFT for $\Omega_{pq} = S_p \wr \fZ_q$. For simplicity we will assume $q$ is prime. 
For the purpose of illustration, let us use (\ref{eq:WreathIdent}) to write down explicitly the partition function of an $S_p \wr \fZ_q$ orbifold theory (for prime $q$) whose connection to R\'enyi entropies we alluded to previously: 
\begin{align}
  Z_{pq,\,S_p \wr \fZ_q}(\tau) &= \frac{1}{q} \left[ Z_{\symq}(\tau)\right]^q + (q-1) \, T_q \left( Z_{\symq}(\tau)\right) \,,
\end{align}
where we used (\ref{eq:genus1Znresult}) for cyclic product orbifold CFTs. In Appendix \ref{sec:wreathCalculation} we carry out a detailed analysis of the spectrum of these partition functions. We find that (as one might expect) such orbifold theories have a universal holographic spectrum provided that $p$ is large enough: 

\begin{theorem}  \label{thm:wreath}
 Given a conformal dimension $\Delta_{max}$ there exists $p_*$ such that the partition function $Z_{S_p\wr \fZ_q}(x)$ is universal for $p > p_*$ up to the order $\mO\left(x^{\Delta_{max}}\right)$, i.e.,
 \begin{align}
   x^{\frac{pqc}{12}} \, Z_{S_p\wr \fZ_q}(x) - x^{\frac{(p-1)qc}{12}} \, Z_{ S_{p-1}\wr \fZ_q}(x) =
   \mO\left(x^{\Delta_{max}}\right) \,,
   \label{eq:thmW}
 \end{align}
with 
\begin{align}
p_* = \max\left\{\frac{16}{c}\Delta_{max},\,\frac{\Delta_{max}}{h_{min}}\right\} \,.
\label{}
\end{align}
\end{theorem}
%

\paragraph{$\fZ_p \wr S_q$ orbifolds:}
\label{sec:WreathSpectra2}
Another obvious example of wreath product theories is ${\cal C}^{\otimes pq}/(\fZ_p \wr S_q)$. Such theories are holographic at large $q$ for reasons which follow almost trivially from Theorem \ref{thm:2}. The proof of the theorem never referred to any explicit properties of the seed theory for the symmetric product orbifold theory apart from a gap in the low-lying spectrum. Thanks to the associativity property of the wreath product, (\ref{eq:WreathIdent}), we conclude that the arguments in the proof of Theorem \ref{thm:2} hold with the only modification that the input theory has central charge $pc$. Therefore, the $\fZ_p \wr S_q$ theories have a spectrum that is independent of $q$ up to $\mO\left(x^{\Delta_{max}}\right)$ if $q\geq \max\left\{\frac{16}{pc}\Delta_{max},\,\frac{\Delta_{max}}{h_{min}}\right\}$. Strictly speaking $h_{min}$ is determined here by the lowest lying states in the $\fZ_p$ cyclic orbifold theory instead of the parent theory ${\cal C}$. However, as we have exemplified in \S\ref{sec:freeb}, this is the same $h_{min}$ as the one determined by the parent theory.

\section{Conclusion}
\label{sec:conclusion}

We have studied spectral properties of central charge $\cc = Nc$ orbifold CFTs $\Cperm = C^{\otimes N}/\ON$ for permutation groups $\ON$ in the large $N$ limit. In general the torus partition function of such a theory is given by a weighted sum over the seed theory partition function evaluated on all unbranched $N$-sheeted covers of the torus consistent with the symmetry group $\ON$. We have shown that at large $N$ this computation drastically simplifies because only the restricted set of untwisted covers dominates. 
Our primary results are the Theorems \ref{thm:0}, \ref{thm:1}, \ref{thm:2} and \ref{thm:oligo} which give the bounds on the spectral density and the constraints on the group $\ON$ such that the resulting theory potentially has a classical string holographic dual.

Generally speaking, such orbifold theories are expected to have a (perhaps stringy) holographic dual if their low lying spectrum is sufficiently sparse. In particular, if the density of low lying states is finite as $N\rightarrow \infty$, then the orbifold CFT has a chance of being holographic. We argued that at the level of group theory this criterion is equivalent to the limiting permutation group $\Omega_\infty$ (if it exists) being oligomorphic. This implies having a finite number of orbits on $\ell$-element sets of natural numbers. For the orbifold CFT to be actually holographic a slightly stronger criterion needs to be satisfied, as has been pointed out in \cite{Hartman:2014oaa}: the density of low lying states must not grow faster than exponential in the energy. While this is the case for many permutation groups with oligomorphic $N\rightarrow\infty$ limit, there exist examples for which the growth rate is faster than exponential. See \cite{Cameron:2000zr} for a number of examples. Theorem \ref{thm:oligo} posits  that for oligomorphic $\Omega_\infty$, whose number of orbits on $\ell$-element sets of natural numbers grows no faster than exponential in $\ell$, one satisfies the criteria guaranteeing a stringy holographic dual.
In particular, we have a bounded growth of low lying states and $F\sim O(1)$ at low temperatures. However, at $T \geq \frac{1}{2\pi}$
the orbifold CFT free energy is dominated at large $N$ by the heavy states that correspond in holography to black hole-like  microstates. This construction gives rise to a large number of permutation orbifold CFTs which should have some dual description in classical string theory. 

We studied two types of permutation orbifolds (and wreath products of them) in greater detail to give an illustration of the connection between the group theoretic constructs and the physical data of the spectral density. On the one hand, for cyclic orbifolds ($\ON = \fZ_N$) the number of low lying states grows with $N$ and accordingly the free energy does not display  features familiar from holography. On the other hand, for symmetric product orbifold theories ($\ON=S_N$) we demonstrated explicitly that the low lying spectrum is non-trivial and dependent on the seed CFT, but also that it is always very sparse (in particular independent of $N$) and therefore washed out in the large $N$ limit (see also \cite{Keller:2011xi}). These two results are made precise in Theorems \ref{thm:1} and \ref{thm:2}.  While these examples are well known, we have found them quite valuable in exemplifying the key features of the abstract group theoretic construction.  

Our analysis was restricted to the case of CFTs on the torus. This led us to study genus one partition functions which encode spectral properties of the CFT. Information about the spectral properties can be used as a diagnostic for whether or not there exists a dual description in classical string theory. However, the spectrum does not encode exhaustive information about the CFT. To make more precise statements about the string theory counterparts, a more detailed knowledge about marginal deformations and higher point correlation functions would provide valuable clues to unearth the dual string theory itself. While it is clear that our general analysis allows for a wide class of holographic orbifold theories, many of them we emphasize should be have very stringy dual descriptions on highly curved AdS${}_3$. It is even plausible that most  states of this theory will never admit a classical supergravity description.\footnote{ We thank E.~Martinec for illuminating discussions on this point.} We take this as an indication that further constraints beyond genus one spectral data need to be formulated in order to demarcate general features of large central charge CFTs with a counterpart in semiclassical gravity.  

For example, the primary (and only) diagnostic we used in our discussion was the sparseness of the low-lying spectrum. To a large extent our results use the general results derived in \cite{Hartman:2014oaa} and employ them to provide constraints on the permutation groups. All this guarantees us is that the CFTs correspond to a classical string theory, perhaps in a string sized \AdS{3} spacetime $\ell_\text{AdS} \sim \ell_s$, with classicality being assured by the large central charge $\cc$ in the $N\to \infty$ limit. One might however wonder if some of these theories could also admit a classical gravitational dual, wherein the stringy degrees of freedom decouple. To attain such a situation, one would have to move away from the free orbifold point which was the focus of  our discussion and attain the gravitational corner of string moduli space. Whether or not this can be done depends on the spectrum of marginal operators of the permutation orbifold $\Cperm$. As far as we are aware, it has never been clearly established, whether even the symmetric orbifold $\Csym$ (without supersymmetry) can be smoothly deformed to attain a classical gravity question. 

A secondary question which does not involve studying deformations away from the free orbifold point, is whether there are other universal features in large central charge CFTs. Apart from the spectral properties, it is known that there are interesting universal properties in the entanglement entropy, such as the vanishing of mutual information for widely separated regions, cf., \cite{Headrick:2010zt,Hartman:2013mia,Asplund:2014coa} for investigations of this issue. The computation of entanglement entropy of course can be mapped to the computation of correlation functions of certain twist operators of the cyclic replica symmetry, or equivalently to the higher genus partition function of the CFT (on branched Riemann surfaces). In fact, the wreath product technology can be employed to the study of R\'enyi entropies (and entanglement entropy) in generic orbifold theories. For orbifold theories that admit a holographic description, it should be possible to verify various phase transitions discovered in the CFT analyses mentioned above (and perhaps used as guide to the string duals by comparing to the gravitational analysis \cite{Faulkner:2013yia,Barrella:2013wja}). 

It would thus be very interesting to study the large $N$ behavior of  the partition functions of 
permutation orbifold CFTs $\Cperm$ on higher genus surfaces.  In general this is a formidable problem; one is required to have the information about the seed CFT partition function on Riemann surfaces of arbitrary genera, 
even to compute the genus-two partition function of $\Cperm$ for arbitrary N. The general group theoretic framework developed by Bantay and described in \S\ref{sec:porbs} can easily be adapted to higher genus 
\cite{Bantay:1998fy,Bantay:2001ay} (we outline the basic constructs in Appendix \ref{sec:HigherGenus}).
In the language of maps from the fundamental group to the permutation group, the main complication is immediately apparent: while unbranched covers of the torus are always (disjoint unions of) tori, covers of a genus $g$ surface generically have even higher genus (by Riemann-Hurwitz). Understanding these surfaces at various points in moduli space makes the analysis quite involved. However, one might hope that at large $N$ such partition functions can be controlled. The reason to hope for this would be, of course, the fact that at genus one we demonstrated how a very restricted set of unbranched covers of the torus dominates the spectrum: most of the complicated twisted sector geometries in the orbifold partition function become irrelevant at large $N$. If true, this would be a fascinating statement about universality of the large central charge permutation orbifolds.\footnote{ We thank Tom Hartman for very useful discussions on this point.}

\acknowledgments 

It is a pleasure to thank Rajesh Gopakumar, Tom Hartman, Veronika Hubeny, Wei Li,  Alex Maloney, Henry Maxfield, 
Shiraz Minwalla, Sameer Murthy  and especially Emil Martinec for enjoyable discussions. 
We would also like to thank the IAS, Princeton for hospitality during the course of this project. 

F.H.\ was supported by a Durham Doctoral Fellowship. M.R.\ was supported in part by the Ambrose Monell foundation,  by the STFC Consolidated Grants ST/J000426/1 and ST/L000407/1,  and by the European Research Council under the European Union's Seventh Framework Programme (FP7/2007-2013), ERC Consolidator Grant Agreement ERC-2013-CoG-615443: SPiN (Symmetry Principles in Nature).

\appendix

\section{Orbifold asymptotics and excited state contributions}
\label{sec:asymproofs}
In this appendix we provide proofs of the statements made in \S\ref{sec:asycs} and \S\ref{sec:wreathpartitions}. The analysis is rather simple and we believe that some of these results are well known by experts. Some of the results in the symmetric orbifold follow from the analysis of \cite{Keller:2011xi}, but our analysis directly deals with the Hecke operators and uses them to extract the relevant asymptotic properties.

\subsection{Dominant contributions at high and low temperatures}
\label{sec:HighAndLowT}

Let us start by examining the behaviour of $Z_{\cyc}$ and $Z_{\sym}$ which are given in 
\eqref{eq:genus1Znresult} and \eqref{eq:GeneralSnresult} respectively,\footnote{ For the $\fZ_N$ orbifold
we focus on prime $N$ for simplicity; the generalization to non-prime $N$ is analogous but more tedious to write out explicitly.}
 as a function of temperature and estimate where the dominant contribution comes form. More specifically, since we have the result for the orbifold partition function in terms of connected covers of tori, we would like to know what covers dominate at a given point in the moduli space (i.e., for fixed $\tau$ in the orbifold theory). This is easy to do in the low and high temperature phases by examining the properties of the Hecke operators' action on the seed partition function $T_M Z(\tau)$. Let us therefore study which terms of the form $Z\left( \frac{M \tau + \kappa \, d}{d^2} \right)$ in each Hecke sum dominate in various regimes.\footnote{ 
 We focus on the thermal contribution alone and work with a rectangular torus: $\tau = i\, \frac{\beta}{2\pi} $ for simplicity, though the generalization to including angular chemical potential with $\tau_1 \neq 0$ is straightforward.}

\begin{itemize}
\item {\bf Low temperatures $\beta \gg 1$:} At low temperatures, the vacuum dominates and the torus partition function behaves as
 \begin{align}
   Z(\tau,\btau) &\sim q^{-c_L/24} \bar{q}^{\,-c_R/24} = \exp\left[\frac{c}{12} \, 2\pi \, \tau_2\right] \,,
    \label{eq:LowTapprox}
 \end{align}
 where here as always we are assuming $c_L = c_R \equiv c$ (note that under this assumption the real part of $\tau$ drops out).  This implies that a term in the Hecke sum behaves as
 \begin{align}
   Z\left( \frac{M \tau + \kappa \, d}{d^2} \right) \sim \exp\left[\frac{c}{12} \frac{M\beta}{d^2}\right] \,.
 \end{align}
 Since the value of $\kappa$ doesn't enter in this asymptotic expression, every allowed value of $\kappa$ gives an equal contribution and the Hecke sum
 can be approximated as follows:
 \begin{align}
   T_MZ(\tau) &= \frac{1}{M} \sum_{d | M} \sum_{\kappa = 0}^{d-1} Z\left(\frac{M\tau + \kappa\, d}{d^2} \right) 
      \sim \frac{1}{M} \sum_{d | M} d \, \exp\left[\frac{c}{12} \frac{M\beta}{d^2}\right]
      \sim \frac{1}{M} \, \exp\left[ \frac{c}{12} M\beta\right] \,,
 \end{align}
 where the last step takes account of the fact that $d=1$ dominates the divisor sum.
 
\item {\bf High temperatures $\beta \ll 1$:} In the high temperature regime, consider first the $\kappa = 0$ term: 
 \begin{align}
   Z\left( \frac{M \tau + 0\cdot d}{d^2} \right) = Z\left( - \frac{d^2}{M \tau}\right) 
             \sim \exp\left[ \frac{ c}{12} \frac{4\pi^2 d^2}{M \beta} \right] \,,
          \label{eq:asympKappaZero}
 \end{align}
 where we used a modular $S$-transformation in the first step and the low temperature expansion in the second step. We claim that all contributions with $\kappa > 0$ are sub-leading in this regime. 
It suffices to show this for $\kappa = 1$; so consider
 \begin{align}
 Z_1\left( \frac{M \tau + 1 \cdot d}{d^2} \right) &= Z_1\left( - \frac{d^2}{M\tau + d}\right) \notag \\
  &= Z_1\left( \frac{\tau \, d^2/M}{|\tau|^2+(d/M)^2} - \frac{d^3/M^2}{|\tau|^2+(d/M)^2}\right) \notag\\
  &\simeq Z_1(M\tau + \text{real}) = Z_1\left( - \frac{2\pi i}{M\beta} \right) \sim \exp\left[ \frac{c}{12} \frac{4\pi^2}{M\beta}\right] \,,
  \label{eq:asympKappaBig}
 \end{align}
 where we used modular invariance ($S$- and $T$-transformations) twice and $|\tau|\ll \frac{d}{M}$ in the third step.  Furthermore, "$\text{real}$" denotes a real part ($\tau$ is purely imaginary) which is irrelevant in the low temperature regime, cf., \eqref{eq:LowTapprox}. The resulting asymptotics in (\ref{eq:asympKappaBig}) is clearly suppressed compared to the contribution of  the $\kappa = 0$ term in \eqref{eq:asympKappaZero}. 
 
 We conclude that the Hecke sum is dominated at high temperatures by the $\kappa=0$ terms:
 \begin{align}
   T_MZ_1(\tau) \sim \frac{1}{M} \sum_{d|M} \exp \left[ \frac{c}{12} \frac{4\pi^2d^2}{M\beta}\right] \sim \frac{1}{M} \exp \left[ \frac{c}{12}
      \frac{4\pi^2 M}{\beta}\right] \,,
 \end{align}
 where the sum over divisors is dominated by the $d=M$ term.
\end{itemize}

We are now in the position to investigate the low and high temperature behavior of $\fZ_N$ and $S_N$ orbifold partition functions. 
Let us start with the $\fZ_N$ orbifold (with $N$ prime). According to the above analysis, the partition function is approximated in 
the respective regimes as follows:
\begin{align}
  Z_{\cyc}(\tau,\btau) &
  	= \frac{1}{N} \left( T_1 Z(\tau,\btau)\right)^N + (N-1) \, T_N Z(\tau,\btau) 
  \notag \\
   &\sim 
  \left\{ \begin{aligned} 
  	    &\frac{1}{N} \exp\left[\frac{c}{12} \beta \right]^N + \frac{(N-1)}{N} \exp\left[\frac{c}{12} N\beta\right]
  	     =\exp\left[\frac{c}{12} N\beta\right]
          &\qquad\;\, (\beta \gg 1 ) \notag\\
      &\frac{1}{N} \exp\left[\frac{c}{12} \frac{4\pi^2}{\beta}\right]^N 
          + \frac{(N-1)}{N}\exp\left[\frac{c}{12} \frac{4\pi^2 N}{\beta}\right]  
         =\exp \left[ \frac{c}{12} \frac{4\pi^2 N}{\beta}\right] 
          &\qquad (\beta \ll 1)\notag  
    \end{aligned}\right.  \notag 
\end{align}
This, of course, agrees with the universal behavior for large central charge CFTs in the low and 
high temperature regimes. The key point here is that the vacuum dominates in either case and we have basically rederived the central result of Cardy in this special case, thereby providing a consistency check of our approximations. 

Let us now do the same for $S_N$ orbifold theories. In that case, the asymptotics are determined by asymptotic Hecke
operators as follows:
\begin{align}
 Z_{\sym}(\tau,\btau)
   &= \sum_{\{m_k\}_N} \prod_{k=1}^N \frac{ (T_k Z(\tau,\btau))^{m_k}}{m_k!} \notag \\
   &\sim \left\{ \begin{aligned}
      & \sum_{\{m_k\}_N} \prod_{k=1}^N \frac{\exp\left[\frac{c}{12} k m_k \beta\right]}{k^{m_k} m_k!} 
      =\exp\left[\frac{c}{12} N \beta \right] 
      &\qquad\;\; (\beta \gg 1) \notag \\
      & \sum_{\{m_k\}_N} \prod_{k=1}^n \frac{\exp\left[\frac{c}{12} \frac{4\pi^2 k m_k}{\beta}\right]}{k^{m_k}m_k!}
        = \exp \left[ \frac{c}{12} \frac{4\pi^2 N}{\beta} \right] 
        & \qquad (\beta \ll 1) \notag
      \end{aligned}\right.  \notag 
\end{align}
where we used $\sum_k km_k = N$ and also the combinatorial fact that $\sum_{\{m_k\}_N} \prod_{k=1}^N \frac{1}{k^{m_k}m_k!} = 1$.
Again, we obtain the correct asymptotics as expected on general grounds.

\subsection{Contribution of excited states}
\label{sec:excitedstates}

Having understood the asymptotic behaviour of the partition function at high and low temperatures let us turn to examining the detailed behaviour of the partition function. The goal is to examine $Z_{\cyc}(\tau,\btau)$ and $Z_{\sym}(\tau,\btau)$ as a function of $N$ and make statements about the limiting behaviour in the 
$N\to\infty$ limit. We first gather some necessary details to extract the contribution of the excited states to the partition sums and then proceed to prove Theorems \ref{thm:1}, \ref{thm:2} and \ref{thm:wreath}.

Let us begin with the character expansion of the seed partition function:
\begin{equation}
Z(\tau,\btau) = q^{-\cn}\,\bq^{-\bcn}
\left[ \chi_{{\mathbb I}}(q)\; \bchi_{{\mathbb I}} (\bq) + 
\sum_{(h,\bh) \in{\cal H}_{\cal C}} \,  \chi_h(q)\; \bchi_{\bh}(\bq) \right] \,,
\label{}
\end{equation}	
where we have isolated the vacuum character for convenience. In terms of the modified eta product 
\eqref{eq:etaM} we have 
\begin{equation}
\chi_{{\mathbb I}}(q) = \prod_{k=2}^\infty \; \frac{1}{1-q^k} = (1-q)\; \eta_1(q)  \,, \qquad \chi_h(q) = \eta_1(q) \; q^h
\label{}
\end{equation}	
Assuming that the states with conformal weights $(h,\bh)$ ($L_0$ and ${\bar L}_0$ eigenvalues) appear with a degeneracy factor  $D_{h,\bh}$ and using the infinite sum representation of the vacuum character in terms of the integer partitions $p_n$ of $n \in {\mathbb Z}$ we can write
\begin{align}
Z(\tau,\bar \tau) &=  q^{-\cn} \, \bq^{-\bcn}\, |\eta_1(q)|^2 
     \left[ (1-q)(1-\bq) + \sum_{(h,\bh) \in{\cal H}_{\cal C}}\; D_{ h,\bh} \; q^h\, \bq^\bh \right]
\nonumber \\
&= q^{-\cn} \, \bq^{-\bcn} \left(\sum_{n,\bn=0}^\infty p_n \, p_\bn \, q^n \, \bq^\bn \right)
     \left[ \sum_{(\Delta,s)\in {\cal \tilde H}_{\cal C}} \tD_{\Delta,s} \, q^{\frac{1}{2}(\Delta+s)} \, \bq^{\frac{1}{2}(\Delta-s)} \right] \,,
\label{eq:Z1q-expansion2}
\end{align}
where $\Delta$ and $s$ take values $(\Delta,s)\in {\cal\tilde H}_{\cal C}
= \{(0,0),(1,1),(1,-1),(2,0)\} \cup \{h+\bh, h-\bh\}_{(h,\bh)\in{\cal H}_{\cal C}}$
to account for the vacuum block and all higher excited states:
\begin{align}
 \tD_{0,0} = \tD_{2,0} = 1 \,, \quad \tD_{1,1} = \tD_{1,-1} = -1 \,,\quad \tD_{h+\bh,\,h-\bh} = D_{h,\bh} \,.
\end{align}

Now we can try to estimate various contributions to the partition function of the permutation orbifold theories.  
For simplicity, let us work with $c=\bc$; generalizations are straightforward.
A basic ingredient that we need for this is the Hecke map. Acting on (\ref{eq:Z1q-expansion2}) it reads
\begin{align}
  T_pZ(\tau,\bar \tau) &= \frac{1}{p} \sum_{d|p} \sum_{\kappa=0}^{d-1} Z\left( \frac{p\tau+\kappa d}{d^2},\frac{p\bar\tau+\kappa d}{d^2}\right)\notag\\
    &= \frac{1}{p} \sum_{d|p} \sum_{\kappa=0}^{d-1} \left[ (q\bq)^{-\frac{c}{24}\frac{p}{d^2}}
       \sum_{n,\bn,\Delta,s} p_n\, p_\bn\, \tD_{\Delta,s}\, q^{\frac{p}{d^2}(n+\frac{1}{2} (\Delta+s))} \, \bq^{\frac{p}{d^2}(\bn +\frac{1}{2} (\Delta-s))} \,
       e^{2\pi i \frac{\kappa}{d} (n-\bn + s)} \right]  \notag\\
    &= \frac{1}{p} \sum_{d|p}  \left[ (q\bq)^{-\frac{c}{24}\frac{p}{d^2}} 
       \sum_{n,\bn,\Delta,s} p_n\, p_\bn\, \tD_{\Delta,s}\, q^{\frac{p}{d^2}(n+\frac{1}{2} (\Delta+s))} \, \bq^{\frac{p}{d^2}(\bn +\frac{1}{2} (\Delta-s))} 
       \, d \,\delta_{d|n-\bn+s} \right] \,,
       \notag \\
      & = \frac{1}{p} \sum_{d|p}  d\,  x^{-\frac{c}{12} \frac{p}{d^2}}  \left[
       \sum_{n,\bn,\Delta,s} p_n\, p_\bn\, \tD_{\Delta,s}\, x^{\frac{p}{d^2}(n+\bn+\Delta)}  
       \, \,\delta_{d|n-\bn+s} \right] \,. 
\end{align}
where $\delta_{d|n-\bn+s}$ comes from summing over $\kappa$; it takes the value $1$ if $d|n-\bn+s$ and $0$ otherwise. In the last line we have further restricted to purely imaginary $\tau$ after accounting for the phases in the Hecke sum.

We can now simplify the action of the $\delta$-function; it gives non-zero contribution in three distinct cases:
$n-\bn+s = m \, d$ with $m<0$, $m>0$ or $m=0$. We want to rewrite the sums over $n$ and $\bn$ in these three cases. 
In the first case ($m<0$), replace $\bn = n+s+|m|d$ and sum over $(n,|m|)$. In the second case ($m>0$), replace $n = \bn - s + m d$ and 
sum over $(\bn, m)$. In the third case, replace $\bn = n+s$ and only sum over $n$. Altogether this yields:
\begin{align}
 T_pZ(x) &= \frac{1}{p} \sum_{d|p}d\, x^{-\frac{c}{12} \frac{p}{d^2}} \sum_{\Delta, s}  \tD_{\Delta,s}  \bigg[ 
    \sum_{m,n=0}^\infty \bigg( p_n \, p_{n+md+s} \, x^{\frac{p}{d^2} (2n+md+\Delta+s)} \notag\\
      &\qquad\qquad\qquad\qquad\qquad  +p_n \, p_{n+md-s}\,  x^{\frac{p}{d^2} (2n+md+\Delta-s)} \bigg)
     + p_n \, p_{n+s} \, x^{\frac{p}{d^2} (2n+\Delta+s)} \bigg] \notag\\
   &= x^{-\frac{cp}{12} } \sum_{d|p} x^{\frac{cp}{12} \left(1-\frac{1}{d^2}\right)} \sum_{\Delta, s} \frac{d}{p} \,  \tD_{\Delta,s}  \bigg[ 
    \sum_{m,n=0}^\infty \bigg( p_n \, p_{n+md+s} \, x^{\frac{p}{d^2} (2n+md+\Delta+s)} \notag\\
      &\qquad\qquad\qquad\qquad\qquad  +p_n \, p_{n+md-s}\,  x^{\frac{p}{d^2} (2n+md+\Delta-s)} \bigg)
     + p_n \, p_{n+s} \, x^{\frac{p}{d^2} (2n+\Delta+s)} \bigg] \,,
 \label{eq:HeckeGeneralExpand}
\end{align}
where we factored out an overall $x^{-\frac{cp}{12} }$ which corresponds to the leading vacuum contribution of the  orbifold theory with central charge $cp$. In the following two subsections we will use the result (\ref{eq:HeckeGeneralExpand}) to make precise statements about
the low-lying spectrum of $\fZ_N$- and $S_N$-orbifold theories at large $N$.

As an illustration of the degeneracies above, consider the free boson at $R=1$ whose partition function is given in \eqref{eq:bosonR1}. In this case the degeneracies can be explicitly computed 
\begin{equation}
\begin{split}
  (\mD_{1i})_i &= (2,\ldots) \,, \qquad\qquad\;\;\,\, (\alpha^1_i)_i = \left(\tfrac{1}{4},\ldots\right)\,,\notag\\
  (\mD_{2i})_i &= (2,4,2,\ldots) \,,\qquad\quad (\alpha^2_i)_i = \left(\tfrac{1}{8},\tfrac{1}{4},\tfrac{1}{2},\ldots\right) \,,\notag\\
  (\mD_{3i})_i &= (3,6,\ldots) \,,\qquad\qquad (\alpha^3_i)_i = \left(\tfrac{2}{9},\tfrac{11}{36},\ldots\right) \,,\notag\\
  (\mD_{4i})_i &= (2,4,8,4,\ldots) \,,\qquad\, (\alpha^4_i)_i = \left(\tfrac{1}{4},\tfrac{15}{48},\tfrac{3}{8},\tfrac{1}{2},\ldots\right) \,,\notag\\
  (\mD_{5i})_i &= (5,10,\ldots) \,,\qquad\quad\;\; (\alpha^5_i)_i = \left( \tfrac{2}{5},\tfrac{9}{20},\ldots\right) \,,\notag\\
  (\mD_{6i})_i &= (2,3,6,\ldots) \,,\qquad\quad\, (\alpha^6_i)_i = \left( \tfrac{3}{8},\tfrac{4}{9},\tfrac{35}{72},\ldots\right) \,,\notag\\
  (\mD_{7i})_i &= (\ldots) \,,\qquad\qquad\qquad (\alpha^7_i)_i = \left( \ldots\right) \,,\notag\\
  (\mD_{8i})_i &= (2,\ldots) \,,\qquad\qquad\quad\, (\alpha^8_i)_i = \left( \tfrac{1}{2},\ldots\right) \,,\notag\\
  \text{for } j\geq 9: \qquad (\mD_{ji})_i &= (\ldots) \,, \qquad\,\qquad\qquad (\alpha^j_i)_i = \left( \ldots \right)   \,, \qquad\qquad\qquad\qquad\notag 
\end{split}
\end{equation}
where we chose the a cutoff $\Delta_{max} = \frac{1}{2}+\eps$, i.e., we only keep track of those terms in the Hecke operators which have powers $\leq\frac{1}{2}$. Hecke operators $T_{j\geq 9}Z(x)$ are approximated as $\frac{1}{j} \, x^{-\frac{cj}{12}}$.

\subsubsection{Excited states in the cyclic orbifold (proof of theorem \ref{thm:1})}
\label{sec:nLargeSpectrumZn}

As we have seen earlier the small-$x$ expansion of $Z_{\cyc}(x)$, is dominated by the vacuum term.  We want 
explicitly find the  leading corrections from the excited states and to derive \eqref{eq:ThmZ}. 

We start the proof by considering the case of $N$ being prime first. In that case we get from Eqs.\ 
\eqref{eq:genus1Znresult} and  \eqref{eq:HeckeGeneralExpand}:
\begin{equation}
Z_{\cyc}(x) =
 \frac{1}{N} Z(x)^N + (N-1)\, x^{-\frac{Nc}{12}} \left\{ \left[ \frac{1}{N} + \mO\left( x^{N h_{min}}\right)\right] 
 + x^{\frac{Nc}{12} \left( 1-\frac{1}{N^2}\right)} \left[ 1 + \mO\left( x^{\frac{1}{N} h_{min}}\right)\right] \right\} \,,
\end{equation}
where the two square brackets come from the divisors $d=1$ and $d=N$ in the Hecke sum $T_N Z(x)$, respectively and $h_{min}$ is the smallest conformal dimension as defined in the statement of 
Theorem \ref{thm:1}. The terms in the curly braces include contributions from both the untwisted (through 
$h_{min}$) and the twisted sector (note the twist operator dimension $\frac{Nc}{12} \left(1-\frac{1}{N^2}\right)$ determining the contribution of the last term). Clearly, all corrections to $1$ in the first square bracket become irrelevant if $N$ is very large. So requiring the  corrections to be at least $\mO(x^{\Delta_{max}})$ if 
$N\geq \frac{\Delta_{max}}{h_{min}}$ for the untwisted sector or 
\begin{align}
  \frac{Nc}{12} \left( 1-\frac{1}{N^2}\right) \geq \Delta_{max} \qquad \Leftrightarrow \qquad  
     N \geq \frac{6 \Delta_{max}}{c}+\sqrt{\left(\frac{6 \Delta_{max}}{c}\right)^2 + 1} \,.
\end{align}
for the twisted sector. Putting these together results in the conditions stated in Theorem \ref{thm:1} for prime $N$. 

Let us now consider the case  where is $N$ non-prime.  The cyclic orbifold partition function is then given by 
\eqref{eq:Z1NonPrime}. The structure of leading contributions in that expression is 
\begin{align}
 Z_{\cyc}(x) &= \frac{1}{N} Z(x)^N + \frac{1}{N} x^{-\frac{Nd}{12}} \bigg[ 
    \sum_{x=1}^{N-1} x^{\frac{cN}{12} \left( 1-\frac{(N,x)^2}{N^2}\right)} \left(1 + \mO\left(x^{\frac{(N,x)}{N}
    h_{min}}\right) \right) \notag\\
    &\qquad\qquad + \sum_{x=1}^{N-1} \left(1 + \mO\left(x^{\frac{N}{(N,x)} h_{min}}\right)\right) 
    + \sum_{x,y=1}^{N-1} x^{\frac{Nc}{12}\left( 1 - \frac{(N,x)^2}{N^2}\right)} \left(1 + 
    \mO\left(x^{\frac{(N,x)^2}{N(N,x,y)}}\right)\right) \bigg]
   \label{eq:CyclicNonPrimeExpand}
\end{align}
The first and the third sum become of order $\mO\left(x^{\Delta_{max}}\right)$ provided that we have  
\begin{align}
  \frac{Nc}{12} \left( 1-\frac{(N,x)^2}{N^2}\right) \geq \Delta_{max} \qquad \forall\, x=1,\ldots,N-1 \,.
\end{align}
The minimum bound on $N$ for which this is satisfied depends, of course, on the number theoretic properties of $N$. However, the worst case that can happen is $N$ being divisible by $2$ whence 
$\text{max}_x \{(N,x)\} = \frac{N}{2}$. 
In that case the above inequality is satisfied for $N \geq \frac{16 \Delta_{max}}{c}$. 
The second sum in (\ref{eq:CyclicNonPrimeExpand}) is more involved. For instance, i
f $N$ is divisible by $2$,  then there will be some sub-leading terms of order $\mO(x^{2\,h_{min}})$. However, such a term would come with a pre-factor  that is independent of $N$. 
Taking into account the overall factor of $\frac{1}{N}$ in \eqref{eq:CyclicNonPrimeExpand}, 
we can thus conclude that such twisted sector states become irrelevant as $N \gg 1$. 
For completeness of the finite-$N$ result, we can, however, still include this sum in the final result. This leads
to the stated result \eqref{eq:ThmZ}.

\subsubsection{Excited states in the symmetric orbifold (proof of theorem \ref{thm:2})}
\label{sec:nLargeSpectrumSn}

Let us now examine the symmetric orbifold theory and see the stabilization of the partition function at large $N$ as stated in \eqref{eq:thmS}. This establishes the universality of the symmetric orbifold partition sum at large $N$.

To construct the symmetric orbifold partition function we need to estimate the contribution from the Hecke operators $T_k$ acting on the seed partition sum, for  $k=1,\ldots,N$. Let us abbreviate their expansion as determined in \eqref{eq:HeckeGeneralExpand} to (this defines  $\mD_{ki}$ and exponents $\alpha_i^k$)
\begin{align}
  T_kZ(x) &= \frac{1}{k} x^{-\frac{ck}{12}} \left( 1 + \sum_i \mD_{ki} \, x^{\frac{c}{12} \alpha_i^k} \right) \,,
  \label{eq:HeckeAbbreviate}
\end{align}
Using (\ref{eq:GeneralSnresult}), we obtain 
\begin{align}
  Z_{\sym}(x) &= x^{-\frac{cN}{12}} \sum_{\{m_k\}_N} \prod_{k=1}^N \frac{1}{k^{m_k}\,m_k!} 
                 \left( 1 + \sum_i \mD_{ki} \, x^{\frac{c}{12} \alpha_i^k}\right)^{m_k}  \notag \\
              &\simeq x^{-\frac{c N}{12}} \sum_{\{m_k\}_N}\left( \prod_{k=1}^N \frac{1}{k^{m_k}\,m_k!}\right)
                 \left[ \sum_{[s^1_I]_{m_1}}\cdots \sum_{[s^N_I]_{m_N}}\prod_{j=1}^N {m_j \choose s_0^j, \cdots, s_N^j}
                 \prod_{i=1}^{\mu_j} (\mD_{ji})^{s_i^j} \, x^{\frac{c}{12} \alpha_i^j \, s_i^j} \right]
         \label{eq:Z1symCalc1}
\end{align}
where $[s_I^j]_{m_j}$ denotes a partition of ${m_j}$, i.e., a set of integers 
$s_0^j,\ldots,s_{\mu_j}^j \in \{0,\ldots,m_j\}$ such that $\sum_{I=0}^{\mu_n} s_I^j = m_j$. So $s_0^j$ counts how many copies of the leading $1$ in the $j$-th Hecke operator (\ref{eq:HeckeAbbreviate}) are being taken
in the multinomial term and similarly $s_i^j$ counts the number of non-trivial terms $\mD_{ji} \, x^{\frac{c}{12} \alpha_i^j}$ with a degeneracy factor
given by the multinomial coefficient\footnote{ Note that we use the uppercase letter $I$ to denote indices including $0$, whereas lower case indices
$i,j$ do not include $0$.}
\begin{align}
  {m_j \choose s_0^j ,\ldots, s_N^j} \equiv \frac{m_j!}{s_0^j! \cdots s_N^j!} \,.
\end{align}
For the $j^{\rm th}$ Hecke operator there is a maximum value for the index, $\mu_j$, 
such that all higher order terms start at $\mO\left( x^{\Delta_{max}}\right)$. This is indicated by
the $\simeq$ symbol for in going from the first to the second line of 
\eqref{eq:Z1symCalc1} we have dropped the higher order terms.
This expression can be simplified further by isolating  the sums over the parameters $s_0^j$ which count the leading order contributions of Hecke operators that are present in each term
and removing them from the partitions $\{m_k\}_N$: 
\begin{align}
 Z_{\sym}(x) &\simeq x^{-\frac{c N}{12}} \sum_{\{m_k\}_N} \sum_{s_0^1=0}^{m_1}\cdots \sum_{s_0^N=0}^{m_N} 
                \left(\prod_{k=1}^N \frac{1}{k^{s_0^k} \, s_0^k!} \right) \left(\prod_{k=1}^N \frac{1}{k^{m_k-s_0^k}}\right) \notag\\
                &\qquad\qquad\qquad\qquad\qquad\qquad\qquad
                \times \left[ \sum_{[s^1_i]_{m_1-s_0^1}}\cdots \sum_{[s^N_i]_{m_N-s_0^N}} \prod_{j=1}^N \prod_{i=1}^{\mu_j}
                   \frac{(\mD_{ji})^{s_i^j}}{s_i^j!} \, x^{\frac{c}{12} \alpha_i^j s^j_i} \right] \,,
          \label{eq:Z1symCalc2}
\end{align}
where $[s^j_i]_{m_j-s_0^j}$ denotes a partition $(s^j_1,\ldots,s^j_{\mu_j})$ of $m_j-s_0^j$.
Written in the form \eqref{eq:Z1symCalc2}, all terms only depend on $m_j-s_0^j$, so the sum over
 $\{m_k\}_N$ is really a sum over $\{m_k\}_{N'}$ with $N' \equiv N - \sum_k \, k s_0^k$ in the following sense:
\begin{align}
  Z_{\sym}(x) &\simeq x^{-\frac{cN}{12}} \sum_{\substack{s_0^1,\ldots,s_0^N = 0 \\  \sum_k ks_0^k \leq N}}^N
                 \left(\prod_{k=1}^N \frac{1}{k^{s_0^k} \, s_0^k!} \right)
                 \sum_{\substack{\{m_k\}_N \\ m_k \geq s_0^k}}  
                 \left(\prod_{k=1}^N \frac{1}{k^{m_k-s_0^k}}\right) \notag\\
                &\qquad\qquad\qquad\qquad\qquad\qquad \times 
                   \left[\sum_{[s^1_i]_{m_1-s_0^1}}\cdots \sum_{[s^N_i]_{m_N-s_0^N}} \prod_{j=1}^N \prod_{i=1}^{\mu_j}
                   \frac{(\mD_{ji})^{s_i^j}}{s_i^j!} \, x^{\frac{c}{12} \alpha_i^j s^j_i} \right] \notag\\
              &= x^{-\frac{cN}{12}} \sum_{\substack{s_0^1,\ldots,s_0^N = 0 \\  \sum_k ks_0^k \leq N}}^N
                 \left(\prod_{k=1}^{N'} \frac{1}{k^{s_0^k} \, s_0^k!} \right)
                 \sum_{\substack{\{m_k\}_{N'} \\ N' \equiv N-\sum_k ks_0^k}}  
                 \left(\prod_{k=1}^{N'} \frac{1}{k^{m_k}}\right) \notag\\
                &\qquad\qquad\qquad\qquad\qquad\qquad \times 
                   \left[\sum_{[s^1_i]_{m_1}}\cdots \sum_{[s^{N'}_i]_{m_N'}} \prod_{j=1}^{N'} \prod_{i=1}^{\mu_j}
                   \frac{(\mD_{ji})^{s_i^j}}{s_i^j!} \, x^{\frac{c}{12} \alpha_i^j s^j_i} \right] \notag\\
              &= x^{-\frac{cN}{12}} \sum_{\sigma=0}^N \sum_{\{s_0^k\}_\sigma} 
                 \left(\prod_{k=1}^\sigma \frac{1}{k^{s_0^k} \, s_0^k!} \right)
                 \sum_{\{m_k\}_{N-\sigma}}
                 \left(\prod_{k=1}^{N-\sigma} \frac{1}{k^{m_k}}\right) \notag\\
                &\qquad\qquad\qquad\qquad\qquad\qquad \times 
                   \left[\sum_{[s^1_i]_{m_1}}\cdots \sum_{[s^{N-\sigma}_i]_{m_{N-\sigma}}} \prod_{j=1}^{N-\sigma} \prod_{i=1}^{\mu_j}
                   \frac{(\mD_{ji})^{s_i^j}}{s_i^j!} \, x^{\frac{c}{12} \alpha_i^j s^j_i} \right]\,,
            \label{eq:Z1symCalc2.5}
\end{align}
where, as before, $\{s_0^k\}_\sigma$ denotes the set of all integer $\sigma$-tuples $(s_0^1, \ldots,s_0^\sigma)$ such that $\sum_k k\, s_0^k = \sigma$. 
By definition, the terms in the second and third line with $\sigma = N$ (i.e., $N'=0$) are just $1$.  
The last equation of (\ref{eq:Z1symCalc2.5}) has the advantage that all the terms which depend on the specifics of the underlying CFT do not depend on $s_0^k$ 
any more. Using $\sum_{\{s_0^k\}_\sigma} \left(\prod_{k=1}^\sigma \frac{1}{k^{s_0^k} \, s_0^k!} \right) = 1$, 
we can therefore factor out every dependence on $\{s_0^k\}_\sigma$: 
\begin{align}
  Z_{\sym}(x) &\simeq x^{-\frac{cN}{12}} \sum_{\sigma=0}^N\sum_{\{m_k\}_{N-\sigma}}
                   \sum_{[s^1_i]_{m_1}}\cdots \sum_{[s^{N-\sigma}_i]_{m_{N-\sigma}}} 
                   \left[ \prod_{j=1}^{N-\sigma} \frac{1}{j^{m_j}} \prod_{i=1}^{\mu_j}
                   \frac{(\mD_{ji})^{s_i^j}}{s_i^j!} \, x^{\frac{c}{12} \alpha_i^j s^j_i} \right] \notag \\
              &= x^{-\frac{cN}{12}} \sum_{\sigma=0}^N
                   \sum_{\substack{\{s_i^j\} \\ \sum_{i,j} j \, s_i^j = N-\sigma}} 
                   \left[ \prod_{j=1}^{N-\sigma}  \prod_{i=1}^{\mu_j}
                   \frac{(\mD_{ji})^{s_i^j}}{j^{s_i^j} \, s_i^j!} \, x^{\frac{c}{12} \alpha_i^j s^j_i} \right] \,,
          \label{eq:Z1symCalc3}
\end{align}
where the last sum runs over all integers $s_i^j$ for $j=1,\ldots,\sigma$ and $i=1,\ldots,\mu_j$ such that $\sum_{i,j} j \, s_i^j = N-\sigma$.

We are now in the position to see the universality of the expression \eqref{eq:Z1symCalc3}. To this end, compare \eqref{eq:Z1symCalc3} for $N$ and $N-1$: one can easily see that the term with index 
$\sigma\neq 0$ in $x^{cN/12} Z_{\sym}(x)$ is the same as
the term with index $\sigma-1$ in $x^{c(N-1)/12} Z_{(N-1),S}(x)$. 
Thus in the difference of these two partition functions only the $\sigma = 0$ term survives:
\begin{align}
  x^{\frac{cN}{12}} \, Z_{\sym}(x) - x^{\frac{c(N-1)}{12}} \, Z_{(N-1),S}(x) 
     = \sum_{\substack{\{s_i^j\} \\ \sum_{i,j} j \, s_i^j = N}} 
       \left[ \prod_{j=1}^{N}  \prod_{i=1}^{\mu_j}
       \frac{(\mD_{ji})^{s_i^j}}{j^{s_i^j} \, s_i^j!} \, x^{\frac{c}{12} \alpha_i^j s^j_i} \right] 
       + \mO\left( x^{\Delta_{max}}\right) \,.
     \label{eq:Z1symCalc4}
\end{align}
We are now only left with the task to estimate a minimum value of $N$ such that the right hand side of this equation is entirely $\mO\left( x^{\Delta_{max}}\right)$. In order to do this, we need the specific form of 
$\alpha_i^j$ which can be obtained from \eqref{eq:HeckeGeneralExpand} and \eqref{eq:HeckeAbbreviate}. 
Clearly it is sufficient to consider the smallest exponents in each Hecke sum since
\begin{align}
 \alpha_i^j \geq \alpha_1^j = j \cdot \min_{1\neq d|j}\left\{\frac{12}{c}\,h_{min}\,,\,\left(1-\frac{1}{d^2}\right)\right\}
  \geq \frac{12}{c}\,j \cdot \min\left\{ h_{min} \, , \, \frac{c}{16} \right\} \equiv \frac{12}{c}\, j \cdot \tilde{h}_{min}
\end{align}
with $h_{min}$ defined in Theorem \ref{thm:1} and $\tilde{h}_{min}$ defined through the above equation. 
We can thus estimate the smallest exponent of $x$ in \eqref{eq:Z1symCalc4} as follows: 
\begin{align}
  \frac{c}{12} \sum_{j=1}^N \sum_{i=1}^{\mu_j} \alpha_i^j s^j_i 
    &\geq \tilde{h}_{min} \sum_{j=1}^N \sum_{i=1}^{\mu_j} j \, s^j_i
    = N\, \tilde{h}_{min}\,. 
    \label{eq:Est1}
\end{align}
We can therefore guarantee that the smallest exponent occurring on the right hand side of
(\ref{eq:Z1symCalc4}) is at least $\mO\left( x^{\Delta_{max}}\right)$ if
\begin{align}
 B\, \tilde{h}_{min} \geq \Delta_{max} \quad \Leftrightarrow \quad 
    N \geq \frac{\Delta_{max}}{\tilde{h}_{min}} = \max\left\{ \frac{\Delta_{max}}{h_{min}} \, , \,
      \frac{16}{c} \, \Delta_{max} \right\} \,.
    \label{eq:Impose1}
\end{align}
as stated in Theorem \ref{thm:2}. 

\subsubsection{Excited states in wreath product orbifolds (proof of Theorem \ref{thm:wreath})}
\label{sec:wreathCalculation}

This subsection contains a proof of Theorem \ref{thm:wreath} based on the results of \S\ref{sec:nLargeSpectrumSn}.
Recall that
\begin{align}
  Z_1^{S_p \wr \fZ_q}(\tau) &= \frac{1}{q} \left[ Z_{\symp}(\tau)\right]^q + (q-1) \, T_q \left( Z_{\symp}(\tau)\right) \notag \\
  &= \frac{1}{q} \left[ Z_{\symp}(\tau)\right]^q + \frac{q-1}{q} \left[ Z_{\symp}(q\tau) + \sum_{\kappa=0}^{q-1} Z_{\symp}\left(\frac{\tau+\kappa}{q}\right)\right] \,.
  \label{eq:WreathSnZq}
\end{align}
Let us consider the three kinds of terms in this expression separately:
\begin{itemize}
\item $\left[ Z_{\symp}(\tau)\right]^q$: We can easily understand the spectrum of this term using Theorem \ref{thm:2}. The latter says that, given a cutoff $\Delta_{max}$, for $n\geq \max\left\{\frac{16}{c}\Delta_{max} , \, \frac{\Delta_{max}}{h_{min}}\right\}$ we have 
\begin{align}
  x^{\frac{cpq}{12}} &\left[ Z_{\symp}(x)\right]^q - x^{\frac{c(p-1)q}{12}} [ Z_{p-1,S}(x)]^q \notag \\
  & = \left[ x^{\frac{c(p-1)}{12}} Z_{p-1,S}(x) + \mO\left(x^{\Delta_{max}}\right) \right]^q-\left[ x^{\frac{c(p-1)}{12}} Z_{p-1,S}(x)\right]^q \notag\\
  &= \mO\left(x^{\Delta_{max}}\right) \,.
\end{align}
This proves the universality of the spectrum arising from the first term in (\ref{eq:WreathSnZq}). 

\item $Z_{\symp}(q\tau)$: The universality of the spectrum of this term can be proven using exactly the same logic as in the proof of Theorem \ref{thm:2}.
The only difference between $Z_{\symp}(q\tau)$ and $Z_{\symp}(\tau)$ is a rescaling of the exponents of $x$ by $q$. The analysis of \S\ref{sec:nLargeSpectrumSn} is therefore still valid with (\ref{eq:Z1symCalc4}) being replaced by 
\begin{align}
  x^{\frac{cpq}{12}} \, Z_{\symp}(q\tau) - x^{\frac{c(p-1)q}{12}} \, Z_{(p-1),S}(q\tau) 
     = \sum_{\substack{\{s_i^j\} \\ \sum_{i,j} j \, s_i^j = p}} 
       \left[ \prod_{j=1}^{p}  \prod_{i=1}^{\mu_j}
       \frac{(\mD_{ji})^{s_i^j}}{j^{s_i^j} \, s_i^j!} \, x^{\frac{cq}{12} \alpha_i^j s^j_i} \right] 
       + \mO\left( x^{\Delta_{max}}\right) \,.
  \label{eq:WreathBullet1}
\end{align}
By the same reasoning as in \S\ref{sec:nLargeSpectrumSn} the previous equation implies that the spectrum of $x^{\frac{cpq}{12}} \, Z_{\symp}(q\tau)$ is independent of $p$ up to some order $\mO\left(x^{\Delta_{max}}\right)$ provided that 
\begin{align}
p \geq \frac{\Delta_{max}}{q \tilde{h}_{min}} = \max\left\{ \frac{\Delta_{max}}{qh_{min}} \, , \,
      \frac{16}{qc} \, \Delta_{max} \right\} \,.
\end{align}
Clearly, the bigger $q$ is, the smaller $p$ needs to be in order to get a universal spectrum up to $\mO\left(x^{\Delta_{max}}\right)$. In fact, if $q$ is large enough, the $S_p$-orbifolding is not even necessary because the corresponding string state is already that of a very long string to begin with. 

\item $Z_{\symp}\left(\frac{\tau+\kappa}{q}\right)$: This term can be dealt with in a very similar way as the previous one. The additional phase factor only changes the coefficients in the expansion of the Hecke operators $T_kZ_1\left(\frac{\tau+\kappa}{q}\right)$ in terms of which $Z_{\symp}\left(\frac{\tau+\kappa}{q}\right)$ is defined. So if we want to reproduce the argument of \S\ref{sec:nLargeSpectrumSn}, we need to change the coefficients $\mD_{ki}$ in the Hecke operator expansion (\ref{eq:HeckeAbbreviate}) and also rescale exponents of $x$ by $\frac{1}{q}$. The analogue of (\ref{eq:HeckeGeneralExpand}) is
\begin{align}
 &T_kZ_1\left(\frac{\tau+\kappa}{q}\right) \notag\\
   &\quad= x^{-\frac{ckq}{12} } \sum_{d|k} x^{\frac{ckq}{12} \left(1-\frac{1}{(qd)^2}\right)} \sum_{\Delta, s} \frac{d}{k} \,  \tD_{\Delta,s}  \bigg[ 
    \sum_{m,n=0}^\infty \bigg( p_n \, p_{n+md+s} \, e^{2\pi i m\frac{\kappa k}{qd}} \, x^{\frac{k}{qd^2} (2n+md+\Delta+s)} \notag\\
      &\qquad  +p_n \, p_{n+md-s}\, e^{-2\pi i m\frac{\kappa k}{qd}} \, x^{\frac{k}{qd^2} (2n+md+\Delta-s)} \bigg)
     + p_n \, p_{n+s} \, x^{\frac{k}{qd^2} (2n+\Delta+s)} \bigg] \notag \\
   &\quad= \frac{1}{k} \, x^{-\frac{ckq}{12} } \left( x^{\frac{ckq}{12}\left(1-\frac{1}{q^2}\right)} + \ldots \right) \,.
 \label{eq:HeckeGeneralExpand2}
\end{align}
We can clearly see that this expression behaves like that of a central charge $c=\frac{k}{q}$ theory, i.e., after expanding it as appropriate for a $c=kq$ theory, the leading term in the bracket is not $1$, but in fact exponentially suppressed. Plugging this expansion of the Hecke operators into the symmetric product orbifold partition function, we find that the leading behavior is
\begin{align}
 x^{\frac{cpq}{12} } \, Z_{\symp}\left(\frac{\tau+\kappa}{q}\right) = x^{\frac{cpq}{12}\left(1-\frac{1}{q^2}\right)} + \ldots \,.
\end{align}
Assuming $q\geq 2$, this expression is zero (and thus $n$-independent) up to $\mO\left(x^{\Delta_{max}}\right)$ provided 
\begin{align}
 \frac{cpq}{12}\left(1-\frac{1}{q^2}\right) \geq \frac{cpq}{16} \geq \Delta_{max} 
 \quad \Leftrightarrow \quad p \geq \frac{16\Delta_{max}}{cq} \,.
\end{align}
This is consistent with the bounds we found in the previous bullet points. 
\end{itemize}

\section{Orbifold CFTs on higher genus Riemann surfaces}
\label{sec:HigherGenus}

\subsection{General formalism}

The paper \cite{Bantay:1998fy} derives the formal answer to the following question: \textit{given the partition functions $Z_g(\uptau)$ for a RCFT $\mC$ on genus $g$ Riemann surfaces, what is 
the partition function $Z_g^{\perm}(\uptau)$ of the permutation orbifold theory $\Cperm = \mathcal{C}^N/\ON$, where $\ON$ is a permutation group of order $N$?}
The partition function on a closed genus $g$ Riemann surface $\Sigma_g$ is parameterized in terms of  
embeddings $\uptau$ of the genus $g$ fundamental group $\Gamma_g$ into the automorphism group of the upper half-plane $\fH$,
\begin{align}
  \uptau: \qquad \Gamma_g \longrightarrow SL(2,R) \,,
\end{align}
such that $\Sigma_g = \fH/ \uptau(\Gamma_g)$. The fundamental group of a genus $g$ Riemann surface is here understood as the free group generated by $2g$ letters modulo a specific commutation relation:
\begin{align}
\Gamma_g  =  \langle a_1, \, b_1, \, \ldots , a_g,\, b_g \; | \; \prod_{i=1}^g\, [a_i, \, b_i] = 1 \, \rangle \,,
 \label{eq:FundGroup}
\end{align}
where $[a,\,b] = a^{-1}\, b^{-1} \, a \, b$ is the commutator. 

The main result of \cite{Bantay:1998fy} reads as follows: 
\begin{align}
Z_g^{\perm}(\uptau) = \frac{1}{|\ON|} \, \sum_{\phi:\, \Gamma_g \rightarrow \ON} \, \prod_{\xi\in \mO(\phi)} Z_{g_\xi}(\uptau_\xi) \,.
 \label{eq:generalResult}
\end{align}
The sum runs over all homomorphisms from the fundamental group $\Gamma_g$ into the subgroup of permutations, $\ON$. The orbits on $X_N = \{1,\ldots,N\}$ under the action of $\phi$ are denoted by 
\begin{align}
\mO(\phi) = \{ \phi(\Gamma_g)\cdot k \; | \;  k = 1,\ldots,N \} \,.
\end{align}
Furthermore, $g_\xi = |\xi|(g-1)+1$ are the genera of the seed theory partition functions and $\uptau_\xi$ defines a genus $g_\xi$ Riemann surface via
restriction of the original coordinate $\uptau$ to the stabilizer subgroup of any element $\xi^* \in \xi$ of the orbit $\xi$: 
\begin{align}
\uptau_\xi \equiv \uptau|_{S_\xi} : \quad S_\xi \equiv \{ x\in \Gamma_{g_\xi} \, | \, \phi(x) \xi^* = \xi^* \,\} \, \longrightarrow SL(2,R) \,.
\end{align}
This defines an embedding $\Gamma_{g_\xi} \rightarrow SL(2,R)$ because of the existence of an isomorphism $S_\xi \cong \Gamma_{g_\xi}$. So $\Sigma_{g_\xi} = \fH/\uptau(S_\xi)$.

The result (\ref{eq:generalResult}) is extremely general. For genus $g=1$, we automatically have $g_\xi = 1$ independent of $\xi$. The fundamental group in this case is $\Gamma_1 = \fZ \oplus \fZ$ and we recover the results of \S\ref{sec:porbs}. Without further specification of the properties of $\Omega$, the formula (\ref{eq:Z1general}) cannot be further simplified. Therefore, in the following subsections we will make these ideas more explicit by restricting to $\ON = \fZ_N$ and $\ON = S_N$.

\subsection{Cyclic orbifolds on genus $g$ Riemann surfaces}
\label{sec:CycHigherGenus}

Let us consider the general formula (\ref{eq:generalResult}) for a higher genus fundamental group $\Gamma_g$. In order to restrict (\ref{eq:generalResult}) to the subgroup $\ON = \fZ_N$, we need to understand all possible homomorphisms 
$\phi : \; \Gamma_g \rightarrow \fZ_N$, where $\Gamma_g$ is the fundamental group (\ref{eq:FundGroup}). Since elements in $\fZ_N$ always commute, 
any such homomorphism is characterized by an arbitrary mapping of the generators of $\Gamma_g$ to elements of $\fZ_N$. Define the homomorphism associated with 
elements $x_1,\ldots,x_{2g}$ by its action on the generators: 
\begin{align}
  \phi_{\{x_1,\ldots,\,x_{2g}\}}(a_i) = x_i, \quad \phi_{\{x_1,\ldots,\,x_{2g}\}}(b_i) = x_{g+i} \qquad (i = 1,\ldots,g) \,. 
\end{align}

Summing over all such homomorphisms 
is therefore the same as summing over $2g$ elements arbitrarily chosen from $\fZ_N$: 
\begin{align}
 Z^{\cyc}_g(\uptau)  = \frac{1}{N} \sum_{x_1,\ldots,\, x_{2g} \in \fZ_N} \;\prod_{\xi \in \mO(x_1,\ldots,\,x_{2g})} Z_{g_\xi}(\uptau_\xi) \,.
 \label{eq:Zggeneral}
\end{align}
Let us consider first the case of prime $N$. As in the genus $1$ case there are only two possible sets of orbits: if all $x_i$ are trivial, i.e. $x_i = N$, then the 
orbits are $\mO(x_1,\ldots,x_{2g}) = \{\{k\}\}_{k = 1,\ldots,N }$, 
whereas if even one of the $x_i \neq N$, the only orbit is $\xi = \{1,\ldots,N\}$. Accordingly, we can write the partition function as
\begin{align}
Z^{\cyc}_g(\uptau)  = \frac{1}{N} Z_g(\uptau)^N + \frac{1}{N} \sum_{\substack{x_1,\ldots,\, x_{2g} \in \fZ_N , \\ \text{not all }=\,N}}  Z_{N(g-1)+1} \left( \uptau\big{|}_{S(x_1,\ldots,\,x_{2g})}\right) \,,
\end{align}
where the stabilizer is the one corresponding to the homomorphism associated with the elements $x_1,\ldots,x_{2g}$:
\begin{align}
 S(x_1,\ldots,x_{2g}) = \{ a\in \Gamma_g \;|\; \phi_{\{x_1,\ldots,\,x_{2g}\}}(a) = e \} \,.
\end{align}

Now consider arbitrary $N$ (possibly non-prime). Again, if all $x_i$ are trivial, $x_i = N$, then the set of orbits is $\mO(x_1,\ldots,\,x_{2g}) = \{\{k\}\}_{k=1,\ldots,\,N}$ and the contribution
to the partition function from this term is just $Z_g(\uptau)^N$. More generally, one can easily see that for a term that corresponds to elements
$x_1,\ldots,\,x_{2g} \in \fZ_N$, the orbits are 
\begin{align}
\mO(x_1,\ldots,\,x_{2g}) = \{\{g^{(N,x_{1},\ldots,\,x_{2g})}\cdot k\}\}_{k=1,\ldots,\,N} \,,
\end{align}
where $(N,x_{1},\ldots,\,x_{2g})$ denotes the greatest common divisor. 
The genus of such a term in the full partition function (\ref{eq:Zggeneral}) is therefore 
$g_\xi = \frac{N}{(N,x_{1},\ldots,\, x_{2g})}(g-1)+1$.
The cyclic orbifold partition function (\ref{eq:Zggeneral}) for arbitrary $N$ can thus be written as
\begin{align}
   Z_g^{\cyc}(\uptau) &= \frac{1}{N} Z_g(\uptau)^N + \frac{1}{N} \sum_{\substack{x_1,\ldots,\, x_{2g} \in \fZ_N , \\ \text{not all }=\,N}} 
      Z_{\frac{N}{(N,x_1,\ldots,\, x_{2g})}(g-1)+1}\left(\uptau\big{|}_{S(x_1,\ldots,\,x_{2g})}\right)^{(N,x_1,\ldots,\, x_{2g})} \,.
\end{align}
Note that the contributing partition functions are evaluated on Riemann surfaces with genera $d(g-1)+1$ with $d$ being divisors of $g$.
The knowledge of the contributing genera is already quite useful. However, for a full understanding of these theories we would need 
a useful parametrization of the moduli space of Riemann surfaces. This requires more work and will be left to a future publication.

\subsection{Symmetric orbifolds on genus $g$ Riemann surfaces}
\label{sec:HigherGenusSN}

Also for $\ON=S_N$ one can find some general formulae which resemble the form of the genus $1$ derivation in \S\ref{sec:symorb}. First we use the fact that the sum over connected covers of the torus can be equivalently understood in terms of a sum over the finite index subgroups of the fundamental group \cite{Bantay:2000eq}. Equation (\ref{eq:generalResult}) implies
\begin{align}
 Z^{\sym}_g(\uptau) = \frac{1}{N!} \sum_{z\in S_N} \;\prod_{\xi \in \mO(z)} \mZ^{|\xi|}_g(\uptau) \,,
 \label{eq:ZSnGeneral}
\end{align}
where we defined
\begin{align}
 \mZ^{(|\xi|)}_g(\uptau)  = \sum_{[\Gamma_g:H]=|\xi|} Z_{g_\xi}\left(\uptau|_H\right) \,. 
 \label{eq:HeckeGeneralized}
\end{align}
This sum runs over all subgroups $H$ of $\Gamma_1$ with finite index $|\xi|$ (up to conjugation) and the argument of the partition function inside this sum is the original embedding of $\Gamma_g$ restricted to the subgroup $H$. (\ref{eq:HeckeGeneralized}) should be thought of as the higher genus generalization of the Hecke operators. Geometrically, it is just a sum over all inequivalent connected covers of the Riemann surface $\Sigma_g$.

Starting from (\ref{eq:ZSnGeneral}), one can follow precisely the same logic as in \S\ref{sec:symorb} to arrive at the following expression:
\begin{align}
 Z^{\sym}_g(\uptau) 
   = \sum_{\{m_k\}_N} \prod_{k=1}^N \; \frac{ 1}{k^{m_k}\,m_k!}\, \left[\mZ^{(k)}_g(\uptau)\right]^{m_k} \,.
\end{align}
As for the higher genus cyclic orbifolds, the data required to compute this expression is given by the seed theory partition function evaluated unbranched covers of $\Sigma_g$, i.e., on Riemann surfaces with genus $g_\xi \in \{g,\ldots,N(g-1)+1\}$ at various points in moduli space. A detailed understanding of this will be left for future work.


\providecommand{\href}[2]{#2}\begingroup\raggedright\endgroup

\end{document}